\documentclass[12pt,preprint]{aastex}

\usepackage{emulateapj5}

\newcommand{\cm}[1]{~cm$^{#1}$}

\newcommand{\e}[1]{$10^{#1}$}
\newcommand{\ee}[1]{$\times 10^{#1}$}
\newcommand{\ergs}{~ergs\,cm$^{-2}$\,s$^{-1}$}
\newcommand{\hi}{H\,{\sc i}}
\newcommand{\hii}{H\,{\sc ii}}
\newcommand{\kms}{~km\,s$^{-1}$}

\newcommand{\msol}{$M_{\odot}$}
\newcommand{\nh}{$N_{\rm H}$}
\newcommand{\oh}{OH(1720~MHz)}
\newcommand{\jyb}{Jy\,beam$^{-1}$}

\begin{document}

\title{Enhanced abundances in three large-diameter mixed-morphology
supernova remnants}

\author{J. S. Lazendic\altaffilmark{1} \& P. O. Slane}
\affil{Harvard-Smithsonian Center for Astrophysics, 60 Garden street,
Cambridge MA 02138}

\altaffiltext{1}{current addres: Center for Space Research, Massachusetts
  Institute of Technology, Cambridge, MA 02139; jasmina@space.mit.edu}

\begin{abstract}

We present an X-ray study of three mixed-morphology  
supernova remnants (SNRs), 
HB~21, CTB~1 and HB~3, using  archival {\em ASCA} and
{\em ROSAT} data. These data are complemented by archival 
{\em Chandra X-ray Observatory} data for CTB~1 and 
{\em XMM-Newton X-ray Observatory} data for HB~3. The spectra from
HB~21 and HB~3 are well-described with a single-temperature thermal 
plasma in ionization equilibrium,
while a two-temperature thermal plasma is found in CTB~1. We found
enhanced abundances in all three SNRs. The elemental abundance of Mg is 
 clearly enhanced in CTB~1, while HB~21 has enhanced abundances of Si
 and S. The situation is not so clear in HB~3 --- the 
plasma in this SNR either has significantly
  enhanced abundances of O, Ne and Mg, or it has marginally enhanced
  abundances of Mg and under-abundant Fe.  We discuss the
  plausibility of mixed-morphology SNR models for the three
SNRs and the presence of enhanced abundances.
We revise a list of MM SNRs and their properties, compare the
   three SNRs studied here with other members of this class, and
   discuss the presence of enhanced elemental abundances in MM SNRs. 
We also report the {\em ASCA} detection of a
compact source in the southern part of HB~3. The
source spectrum is consistent with a power law with a photon
index of $\sim 2.7$, and an unabsorbed X-ray flux  
of $\sim$\ee{-12}\ergs\ in the 0.5--10.0~keV
band. The column density towards this source differs from 
that towards the SNR, and it is therefore unlikely they are related. 

\end{abstract}


\section{INTRODUCTION}

Mixed-morphology (MM) or thermal composite supernova remnants (SNRs)
are  characterized by a shell-like morphology in the radio band and centrally
peaked thermal emission in the X-ray band
\citep{seward90,jones98,rho98}. Despite the relatively large number  
($\sim 10$\%) of Galactic SNRs that display mixed morphology, a
mechanism that is responsible for producing such a morphology has as yet
not been uniquely identified. There are a few models able to produce
centrally enhanced X-ray emission, such as evaporation of clouds which
are left relatively intact after the passage of the SNR blast wave
\citep[e.g.,][]{white91}, thermal conduction effects  in the remnant
interior  \citep[e.g.,][]{cox99}, or evolution in a medium with a
density gradient viewed along the line of sight \citep{petruk01}. The
X-ray properties of MM remnants have been defined almost exclusively
using {\em
ROSAT} data \citep{rho98}; these SNRs have: 1) centrally peaked thermal X-ray
emission, with or without a fainter shell, 2) a flat radial temperature 
profile, and 3) X-ray emission that arises primarily from shocked interstellar
material and not from ejecta. However, recent studies with {\em ASCA}
indicate that these SNRs have a complex plasma structure, with  
multiple components  \citep[e.g.,][]{rho02} and enhanced
abundances  \citep[e.g.,][]{yamauchi99,slane02}.   
 We have, therefore, searched the  X-ray archives for unpublished 
{\em ASCA} data on MM SNRs, in order study their spectral
properties in more detail and construct a more complete picture of MM SNRs. 
Only three MM SNRs were found with unpublished 
{\em ASCA} data: HB~21, CTB~1 and HB~3, all of which have low 
X-ray brightness emission and, consequently, low number of counts 
($\sim 2000$ in total).  We also used archival {\em Chandra
X-ray Observatory} data of CTB~1, and {\em XMM-Newton X-ray
  Observatory} data of HB~3; HB~21 has not been observed by either of
the two observatories.

HB~21 (G89.0+4.7) has an angular extent of $120\arcmin\ \times
90\arcmin$ and a typical  radio spectral
index of $-0.4$ \citep[e.g.][]{willis73,reich03}.  Optical emission 
has been detected towards the SNR, but it is not clear whether
it is associated with the remnant
\citep{willis73}. The distance estimate of 0.8~kpc to the SNR assumes 
 association with the molecular clouds in the Cyg~OB7 complex
\citep{humphreys78}. CO and \hi\ observations towards the SNR suggest
 an interaction with the ambient medium \citep{tatematsu90,koo01}. X-ray
emission from HB~21 was first detected by the {\em Einstein IPC}
\citep{leahy87,seward90} and studied in more detail with {\em
ROSAT} All-Sky Survey data \citep{leahy96}. 

CTB~1 (G116.9+0.2) is a $\sim 30\arcmin$ diameter SNR with a break-out  
morphology in the northeast region. The radio spectral index of $-0.6$ 
\citep{willis73} and
peak polarization of $\sim 40\%$ at 6~cm \citep{angerhofer77}
confirms the nonthermal nature of the SNR. Optical emission has been
detected from the SNR, with a morphology resembling that in radio
\citep[e.g.,][]{fesen97}.  The distance to  the SNR is uncertain;
suggested values range from 1.5 to 4~kpc  \citep[see
e.g.,][]{fesen97}.  The remnant was first
detected in X-rays by {\em ROSAT} All-Sky Survey and studied in more
detail with the {\em ROSAT} PSPC data \citep{craig97}.

SNR HB~3 (G132.7+1.3) is $90\arcmin \times  120\arcmin$ in diameter 
 and has radio spectral index of $-0.64$
 \citep{landecker87,fesen95,reich03}.  The remnant is located next to
 the  W3 \hii\ complex, and their association, assumed due to their
 proximity,  implies a distance of 2.2~kpc
 to the SNR \citep{routledge91}. Diffuse and filamentary optical
 emission has been detected from the SNR, with the strongest emission
 along the western SNR shell \citep{fesen95}. Studies in the X-ray band
 have been performed  using {\em Einstein} observations
 \citep{leahy85,seward90}. 

 No associated radio pulsar has been found in any of the three SNRs so
far \citep{lorimer98}.  A radio pulsar PSR~J0215+6218 has been
 discovered within (in projection) the HB~3 boundary, but it appears to
 be much older than the remnant and therefore not associated with the SNR
\citep{lorimer98}. Also, an unassociated X-ray pulsar has been discovered
 at the edge of the north-east region of CTB~1 \citep{hailey95}.

In section 2 we describe the data used in this study. In  section 3
we describe results on individual remnants, and compare our  results
with observations at other wavelengths. In section 4 we compare our
results with models for MM SNRs, and discuss the presence of  enhanced
abundances in MM SNRs. The conclusions are given in section 5.


\section{DATA}

\subsection{ASCA data}

We obtained {\em ASCA} data provided with the standard screening process 
 from the HEASARC (High Energy Astrophysics Science
Archive Research Center) public database.  {\em ASCA} \citep{tanaka94}
comprises two detector  pairs: the Gas Imaging Spectrometers (GIS2
and GIS3), each with a field of view 40\arcmin\ in diameter, and  the
Solid-State Imaging Spectrometers (SIS0 and SIS1) with four  CCDs,
each of $11\arcmin \times 11\arcmin$ in size. The SIS(GIS) detectors
cover  an energy band between 0.4(0.6) and 12~keV, have spectral
resolution of about  0.1(0.2)~keV at 1~keV and combined with the X-ray
  telescope provide on-axis angular
 resolution of about 2\arcmin\ (SIS) and 3\arcmin\ (GIS). 

Details of the {\em ASCA} observations of the three SNRs,  including
observational sequence numbers, date of observations,  total exposure
times and observing modes, are listed in Table~\ref{tab-data-asca}.
 A different number of the SIS CCDs can be exposed at a time, giving
 1-, 2-, or 4-CCD modes. 
Each SNR has been observed with at least two pointings.  We used  
standard routines for spectral analysis of {\em ASCA} data 
in Xselect from Ftools analysis package. The data for each SNR 
were taken later in the mission and
 additional processing of the SIS data was  needed 
to correct for degradation of the  CCDs. 
 Thus, we corrected the SIS data for charge transfer inefficiency
(CTI), residual dark distribution (RDD) and dark frame error (DFE) effects
\citep[for more details see e.g.,][]{dotani97}. However,
 even after these corrections, some discrepancy between 
GIS and SIS spectra below 1~keV can
remain\footnote{http://heasarc.gsfc.nasa.gov/docs/asca/watchout.html},  
 potentially leading to 
large discrepancies between SIS- and 
GIS-derived column densities. Therefore, we perform a joint
 spectral fitting of GIS and SIS spectra for the regions common to 
both GIS and SIS detectors, which we refer to in the text as SIS regions. 
We consider separately the entire portions of the remnants that are 
covered in each GIS detector, which we refer to in the text 
as GIS regions.  In pointings where strong point sources
were present, these were removed before extracting the spectra. 

 For background subtraction we used blank-sky observations from
 high Galactic latitude fields from which bright point-like sources
 were removed. Additional  background X-ray
 emission attributed to the Galactic ridge was found to be present, mostly in
 the high-energy ($\ge$3~keV) part of the spectrum,  which was 
 accounted for in our spectral fitting in a following manner. We
 subtracted a blank sky background from a spectrum created using a 
source-free region on the GIS detector. The residuals from this
 subtraction were then fitted with a model consisting of a power law;
 some indication of lines were present but they were too weak to be
 fitted by Gaussians. This was then added as a background model 
in spectral fits, normalized by the solid angle of the source region.

\subsection{ROSAT data}

We obtained the {\em ROSAT} data from the Max-Planck
Institut f\"{u}r Extraterrestrische Physik public
database\footnote{http://www.xray.mpe.mpg.de/rosat/archive/index.html}
and used them primarily to produce X-ray images of the whole
remnants. The Position-Sensitive Proportional Counter
\citep[PSPC;][]{pfeffermann86} on board {\em ROSAT} \citep{trumper83}
has field of view of $\approx 2$\degr, spectral resolution of about
0.45~keV at 1~keV,  energy range of 0.1--2.4~keV,
and angular resolution of 15\arcsec. 

All three SNRs have extents larger then 40\arcmin', the diameter of the central 
PSPC region, where the PSPC response is most
uniform and sensitive. Therefore, all three sources have been observed
with two or more pointings. Observational sequence numbers, dates of
observation and total exposure times are listed in
Table~\ref{tab-data-rosat}.  For creating mosaic images from {\em
ROSAT} PSPC data we used standard procedures in the EXSAS software
package \citep{zimmermann94}. 

We also created PSPC spectra to use in conjunction with
{\em ASCA} data to constrain the column density (since PSPC data
extend to lower energies than the {\em ASCA} detectors). For spectral
analysis of PSPC data, standard routines in Xselect 
 were used. Point sources
were removed when creating {\em ROSAT} spectra. In all cases a
spectrum  for background subtraction was obtained from emission-free
regions at the edge of the detector.  

\subsection{Chandra data}

The observations of CTB~1 were obtained from
  the {\em Chandra} Data Archive. The Advanced Camera for Imaging and Spectroscopy (ACIS) 
 on-board {\em Chandra} consists of two CCD arrays, 
 the ACIS-I comprising of four CCDs arranged in a square, and ACIS-S
 comprising of six linearly adjacent CCDs (ACIS-S).  
Each CCD is about $8\arcmin\times8\arcmin$ in size. Only six CCDs can be used at one time. 
 ACIS covers the energy band between 0.1 and 10~keV and has an energy
 resolution of 0.1~keV at 1~keV, and sub-arcsecond angular resolution.

CTB~1 was observed with an ACIS exposure of 50~ks on 2002 September 14 (ObsID 2810). 
The observations were motivated by a search for a neutron star
associated with the  SNR \citep[e.g.,][]{kaplan04}. Four ACIS-I and
two ACIS-S (S2 and S3) CCDs were used. Most of the SNR was
 covered with the four ACIS-I chips.  Data were taken in full-frame
timed-exposure mode with the standard integration time of 3.2~s.  
 Data were reduced using standard threads in CIAO v.3.1, and
calibration database CALDB v.2.28. The corrections
for the effect of increased CTI and degradation of the low-energy 
(E $<$ 1~keV) quantum efficiency of the ACIS were included in the
standard procedures for CIAO v.3.1.
 The effective exposure time after data processing and screening for
 ``flaring'' pixels was unchanged.
We used weighted response matrices for spectral fitting 
to account for CTI effects across the ACIS CCDs. 
For background subtraction we used the blank-sky observation files, following the
 standard procedure in CIAO v.3.1.

\subsection{XMM-Newton data}

We obtained archival {\em XMM-Newton} data for HB~3 from the HEASARC archive. The
{\em XMM} European Photon Imaging Camera (EPIC)  comprises three
detectors: one pn and two MOS cameras.  They cover the energy band
between 0.2 and 12~keV, and have energy resolution of 0.15~keV at
1~keV. Their on-axis angular resolution is around 6\arcsec\ FWHM and
15\arcsec\ HPD, and the field of view is around 30\arcmin. 

The {\em XMM} observation of HB~3 was carried out on 2003 February 1, with
an 8~ks pn exposure and 3~ks MOS exposure (ObsID 0150960401).  As for
CTB~1, these observations were motivated by a search for a neutron
star associated with the  SNR \citep[e.g.,][]{kaplan04}.  The EPIC-pn
camera was operated in Full Frame mode with a time resolution of 73~ms,
while the two EPIC-MOS cameras were operated in Full Frame mode with a
time resolution of 2.6~s. The thin filter was used for all three cameras.  The
EPIC data reduction was performed using the SAS software package
version 6.0.0.  The event files used for analysis  were created from
observational data files (ODFs)  using the SAS tasks EPCHAIN and
EMCHAIN.  The observations were found to be affected by periods of
high particle background, and the net exposure time after filtering
event files for good time intervals was 3~ks in the pn camera and
1.6~ks in each MOS camera.

In spectral analysis the blank sky observations were used for
background subtraction. We used scripts supplied by the University of 
Birmingham {\em XMM} group\footnote{http://www.sr.bham.ac.uk/xmm2/}
to adjust the blank sky event file to match our data event
file and to extract the source and the background regions 
\citep[for more detail description of procedure see e.g.,][]{sasaki04}. 
We then fitted the source spectrum and the
background spectrum simultaneously, including recommended spectral
components for the {\em XMM} detector background \citep[see e.g.,][]{lumb02}.

\subsection{Radio}

For a comparison with radio data we used the 92~cm observations
from the Westerbork Northern Sky Survey (WENSS), obtained from the
SkyView public database. We smoothed the radio data from their
original resolution of $54\arcsec \times 54\arcsec$\,cosec($\delta$) to
2\arcmin--3\arcmin\ in order to improve signal-to-noise in the images.

\section{RESULTS}

 To examine the SNR morphology we used both PSPC and GIS images. We
generated exposure-corrected PSPC images using photons with energies 
between 0.2 and 2~keV. The GIS images were exposure-corrected, 
background subtracted and masked in regions where the exposure falls below
15\%\ of the maximum value. Photons with energies between 0.7 and 10.0~keV
were used for making the GIS images.

The final spectral analysis was performed using the {\em ASCA} data alone,
 because adding the {\em ROSAT} PSPC spectra did not improve the fit
 or constrain the  parameters derived from spectral fitting significantly.
 We used a model for  interstellar absorption by \citet{morrison83}, in
 combination with a model  for an optically thin thermal plasma in
 collisional ionization equilibrium \citep[CIE; VRAY(MOND);][]{raymond77}.
  We also used a plane-parallel time-dependent ionizing  plasma model
 \citep[VPSHOCK;][]{borkowski01}, and an ionizing collisional plasma
 model which assumes a single ionization parameter \citep[VNEI;][]{mazzotta98}.
  The VPSHOCK model can provide useful information on emission from the SNR boundary
 regions (e.g., NE region in CTB~1 or N3 region in HB~3) where
 we can expect a range of ionization timescales if the evolving
 shock (i.e. forward shock) is present. The central SNR regions 
 were presumably shocked some time ago, and  
 equilibrium models such as VRAY or VNEI are more appropriate.  
 The GIS data can suffer from gain variations, and we
 applied gain corrections when fitting GIS data resulting in gain
 adjustments of 1\%--3\%; these corrections did not always improve
 the fit, nor did they change the fit parameters significantly {and are, therefore, 
not used in the final fits}. 
All the spectra were grouped to ensure a minimum of 25 counts per bin. 
 Due to the low X-ray surface brightness of the SNRs, a detailed spatially
 resolved spectral analysis for each pointing was not
 possible. However, since {\em ASCA} pointings cover different regions
 of the SNRs, they do provide spatially resolved spectral analysis to a limited extent.

\subsection{HB~21 (G89.0+4.7)}

\subsubsection{Morphology}

\begin{figure}
\centering
\includegraphics[height=12cm]{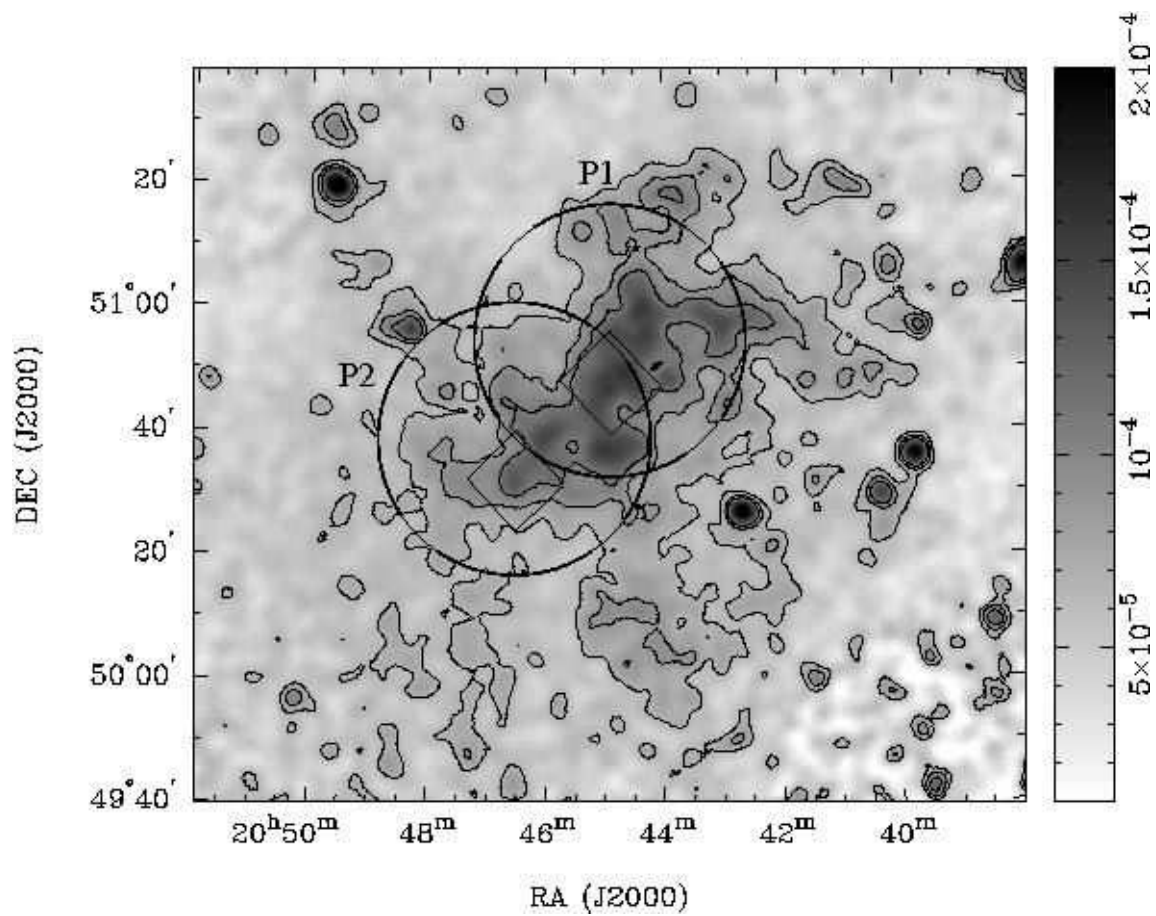}
\caption{Exposure-corrected {\em ROSAT} PSPC greyscale and contour mosaic image of the
 SNR HB~21. The image has been smoothed with a 5\arcsec\ FWHM Gaussian and 
units are in counts~arcmin$^{-2}$~s$^{-1}$. The regions covered in two
 separate pointings (P1 and P2) with {\em ASCA} GIS
 detectors are marked with circles, and squares mark SIS detectors. }
\label{fig-image-hb21}
\end{figure}

\begin{figure}
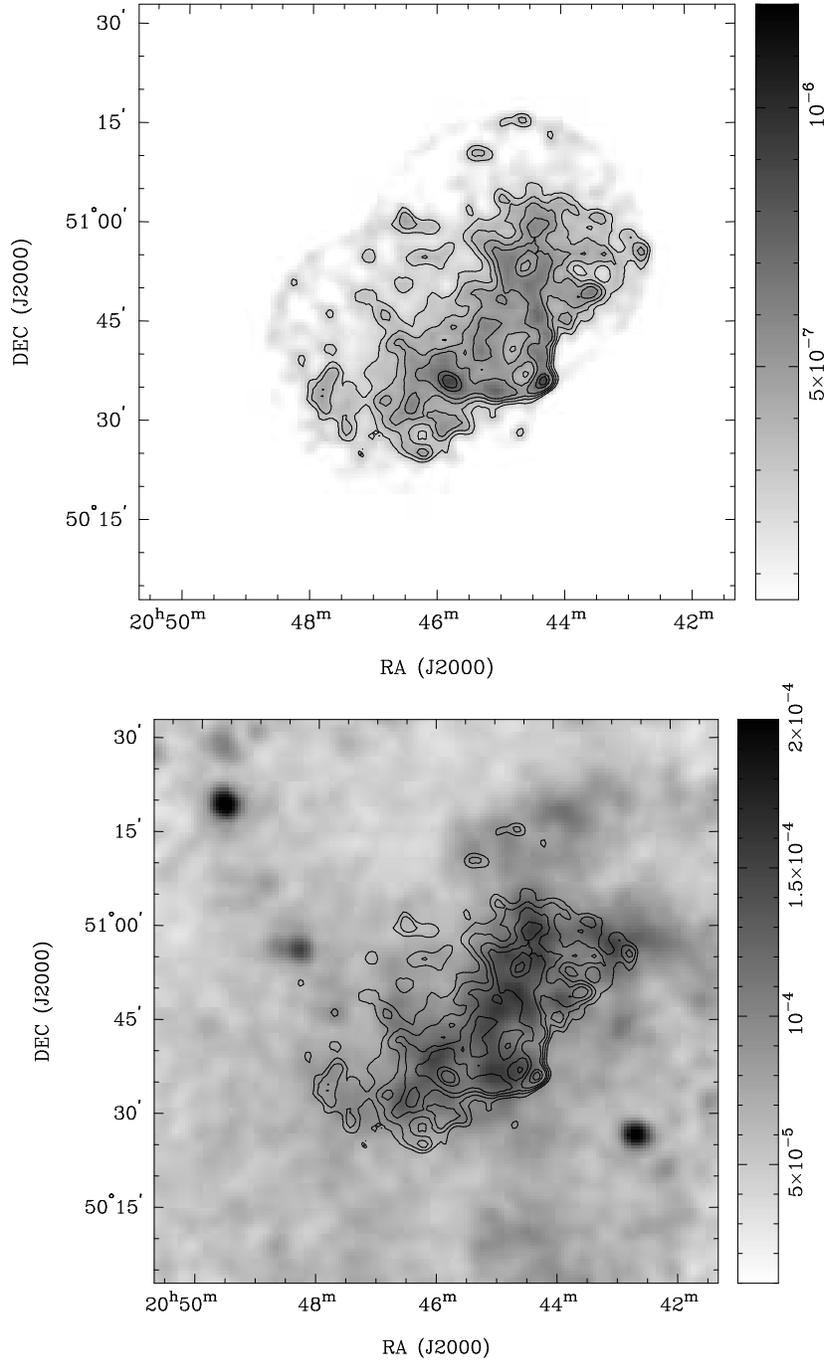

\centering
\includegraphics[height=9cm]{f2a.ps}
\includegraphics[height=9cm]{f2b.ps}
\caption{ {\em Top:} Exposure-corrected and background {\em ASCA} GIS mosaic image of the central region of HB~21. The image has been smoothed with a 2\arcsec\ FWHM Gaussian and 
units are in counts~arcmin$^{-2}$~s$^{-1}$. {\em Bottom:} GIS contours overlaid onto the {\em
ROSAT} greyscale image from Fig.~\ref{fig-image-hb21}. Contour levels are:
30, 40, 50, 60, 80, 90 $\times$ 2\ee{-8} counts~arcmin$^{-2}$~s$^{-1}$.}
\label{fig-gis-hb21}
\end{figure}

Figure~\ref{fig-image-hb21} shows the {\em ROSAT} PSPC mosaic image 
of HB~21, composed from six pointings and smoothed
with a Gaussian of 5\arcsec\ FWHM.   Regions covered by the
{\em ASCA} detectors are also marked. As shown before \citep{leahy96},
the  X-ray emission is centrally concentrated, and appears elongated in
 the northwest-southeast direction.  The {\em ASCA} GIS image of the
central part of the SNR, and the  overlay of the GIS contours onto the
{\em ROSAT} image are shown in Figure~\ref{fig-gis-hb21}.  There is a
good correlation between the two images. The X-ray emission from the
SNR appears clumpy both in the PSPC and GIS images. The GIS image
appears to have a half-circular shell-like feature oriented
northeast-southwest and stretching through
the center which is not obvious in the PSPC image. 
  We note that this shell coincides with the edge of one of the
   GIS (P1) pointings, and thus could be an artifact of the mosaicing.
  To investigate this structure in more detail, we produced an exposure-corrected image of  pointing P2 alone. The half-circular feature is still present, which suggests it is real.
The two brightest clumps of X-ray emission are located along this
bright shell in the GIS image and they do not have corresponding bright
emission in the PSPC image. This suggests that the bright shell and the 
two bright clumps, have the harder spectrum than the 
other parts of the SNR that have corresponding PSPC emission.   
  We also produced soft (0.7--2.5~keV) 
and hard (2.5--10.0~keV) band GIS images, but we found no significant differences, probably due to the relatively low number of counts available.

\begin{figure}
\centering
\includegraphics[height=10cm]{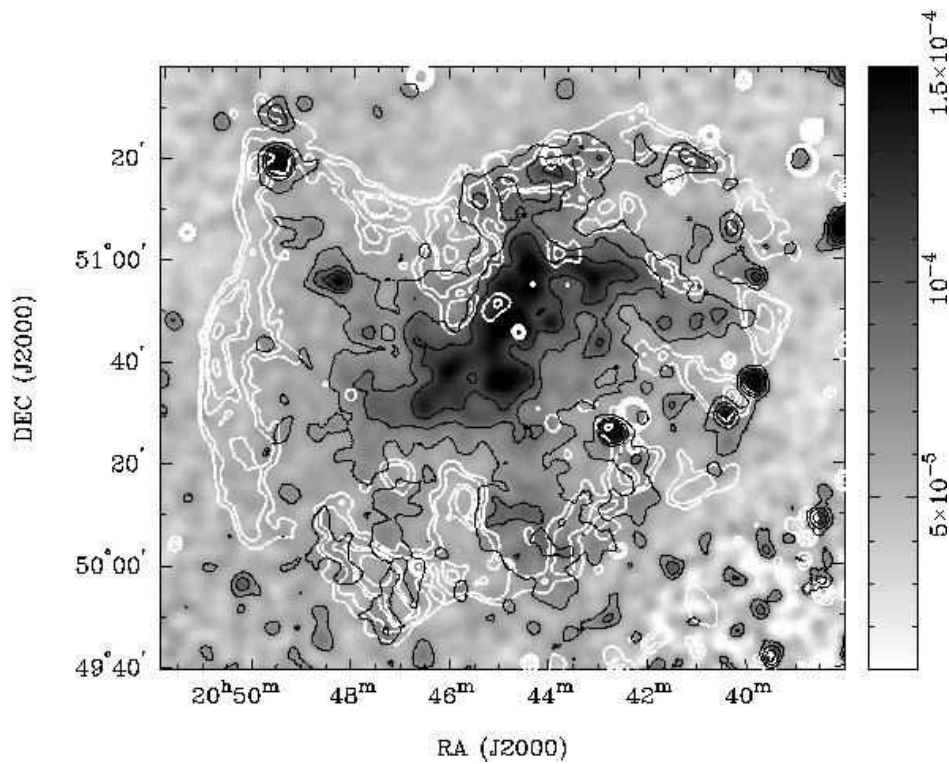}
\caption{{\em ROSAT} PSPC image of HB~21 (greyscale and light
 black contours) overlaid with the 92~cm WENSS radio image 
(white contours), which has been 
smoothed to 2\arcmin\ resolution. Contour levels for the {\em ROSAT}
 image are same as in in Fig.~\ref{fig-image-hb21}, and radio (white)
 contours are: 4, 8, 16, 20, 30, 50, 70, 90 $\times$ 0.01 \jyb . There is no radio
emission corresponding the central X-ray emission. Faint diffuse
emission is seen extending to the northern, southern and western SNR limb.}
\label{fig-radio-hb21}
\end{figure}

Radio and X-ray emission from HB~21 are compared in 
Figure~\ref{fig-radio-hb21},
which shows that there is no radio emission corresponding to the
central X-ray  emission. The fainter, diffuse X-ray emission 
extends towards the radio limbs in all directions except for the east.
\citet{leahy96} suggested that the emission from the eastern  SNR shell is
affected by  absorption from the molecular clouds associated with
HB~21, which have column densities of (2.5--5.5)\ee{21}\cm{-2}
\citep{tatematsu90}. Optical images also imply that extinction towards
the eastern SNR region is higher than elsewhere \citep{willis73}.
\citet{tatematsu90} suggested an interaction between the \hi\ and CO gas
along the eastern SNR shell, on the basis of morphology.  Observations
with improved spatial resolution by \citet{koo01} confirmed the
presence of molecular gas along the eastern SNR boundary, but found
little correlation between features in the molecular cloud and the SNR
brightness distribution, or small-scale morphology.  They also found no
direct kinematic evidence of interaction (e.g., broad molecular lines)
between the eastern clouds and the SNR shell. Instead, broad
  molecular lines are found toward the bright radio regions in the  
 north and south part of the SNR shell. The IR emission is
present along the SNR boundary, and shows large-scale correlation with
CO emission. Using the 60 and 100~\micron\ ratio, \citet{koo01} found
additional support that interaction at the eastern SNR shell is
unlikely, while the IR emission from the north and south clouds has
 a higher 60/100~\micron\ ratio, suggestive of shock origin. The northern
and southern bright radio regions lack X-ray emission, as shown 
in Figure~\ref{fig-radio-hb21}, indicating that X-ray emission from the
SNR has been absorbed or the shock could have gone radiative due to
interaction with molecular clouds.

\subsubsection{Spectra}

\begin{figure}
\centering
\includegraphics[height=8cm]{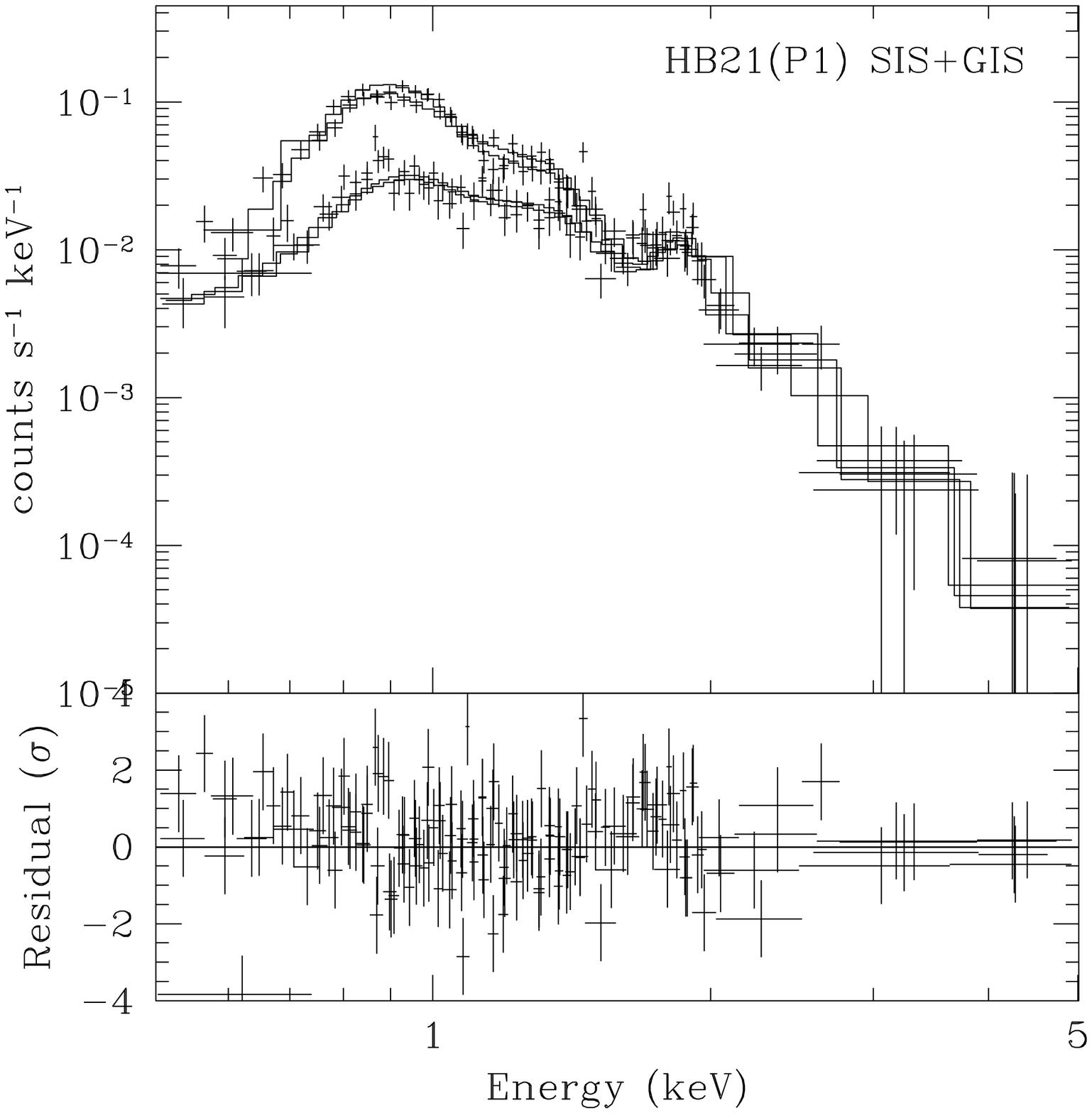}
\includegraphics[height=8cm]{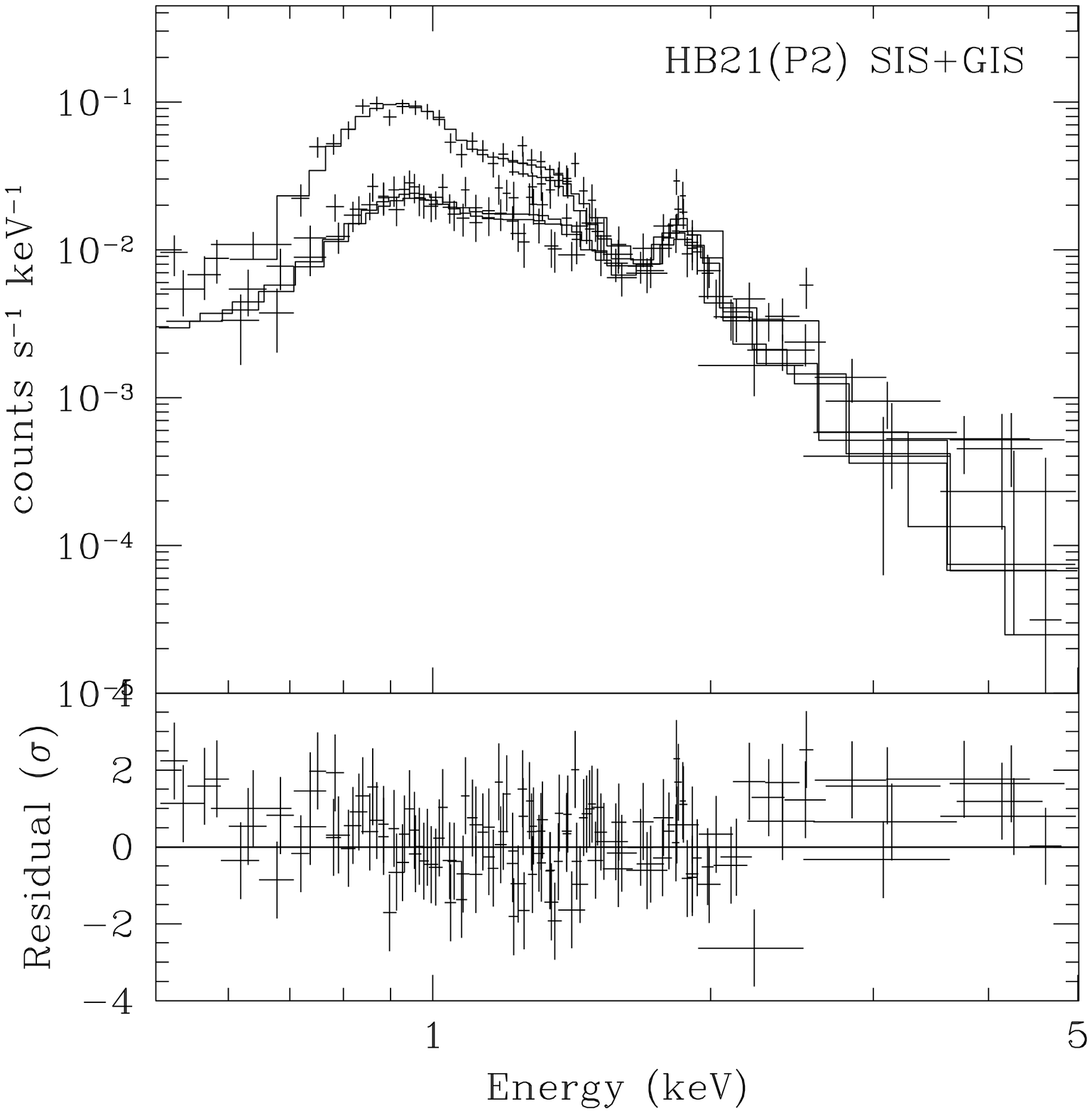}
\caption{GIS and SIS spectra of the SNR HB~21 from the two {\em ASCA}
pointings  marked in Figure~\ref{fig-image-hb21},  common to GIS and SIS detectors, and residuals from the best-fit models.
 In all spectra the Si line is prominent.
Fit parameters are listed in Table~\ref{tab-fit-hb21}.}
\label{fig-spectrum-hb21}
\end{figure}

The joint fits to the GIS and SIS spectra from the two SIS regions of 
HB~21 are shown in Figure~\ref{fig-spectrum-hb21}. 
 The He-like Si line feature is prominent in both the GIS and SIS data. The
 VRAY model gave a satisfactory fit to the spectra, but the 
VNEI model  gave an improved fit ($4\sigma$ significance) in most cases; the VNEI fit
parameters are summarized in Table~\ref{tab-fit-hb21}.  We note that high values of 
$n_e t$ imply that plasma is close to, or have reached equilibrium, and thus, VRAY model is acceptable as well. The  fit with solar 
abundances was improved significantly for some regions
 when the Si and S abundance was thawed.
There is no significant variation in temperature between the two
pointings, but there could be some variation in the column density,
with the P2 pointing having higher column density than the P1 pointing. 
 Slightly enhanced abundances imply that the X-ray plasma in 
HB~21 comprises supernova ejecta mixed in
with the swept up shocked material. The bright shell-like feature revealed
 in the GIS image could be then formed by the reverse shock.

Spectral parameters obtained from the {\em ASCA} data are
broadly in agreement with the fit parameters derived from the previous
study using {\em ROSAT} data alone \citep{leahy96}. In general, column densities inferred from {\em ASCA} spectra tend to be higher than those from the PSPC.  This is 
a direct result of the different bandpasses. The column density
 of $\approx 4$\ee{21}\cm{-2} derived here seems high for an SNR
   as close as 0.8~kpc, but this column density is comparable to the column
density of the eastern molecular clouds observed by
\citet{tatematsu90}. Thus, it seems that most of the column
  density to the SNR is coming from the eastern molecular cloud, and it is
plausible that the eastern  SNR shell could be more absorbed 
than the rest of the SNR. Another possibility is that the shock has encountered a nearly circular (in projection) cavity edge in the southern half of the remnant (as in the Cygnus Loop SNR).


\subsection{CTB 1 (G116.9+0.2)}

\subsubsection{Morphology}

\begin{figure}
\centering
\includegraphics[height=12cm]{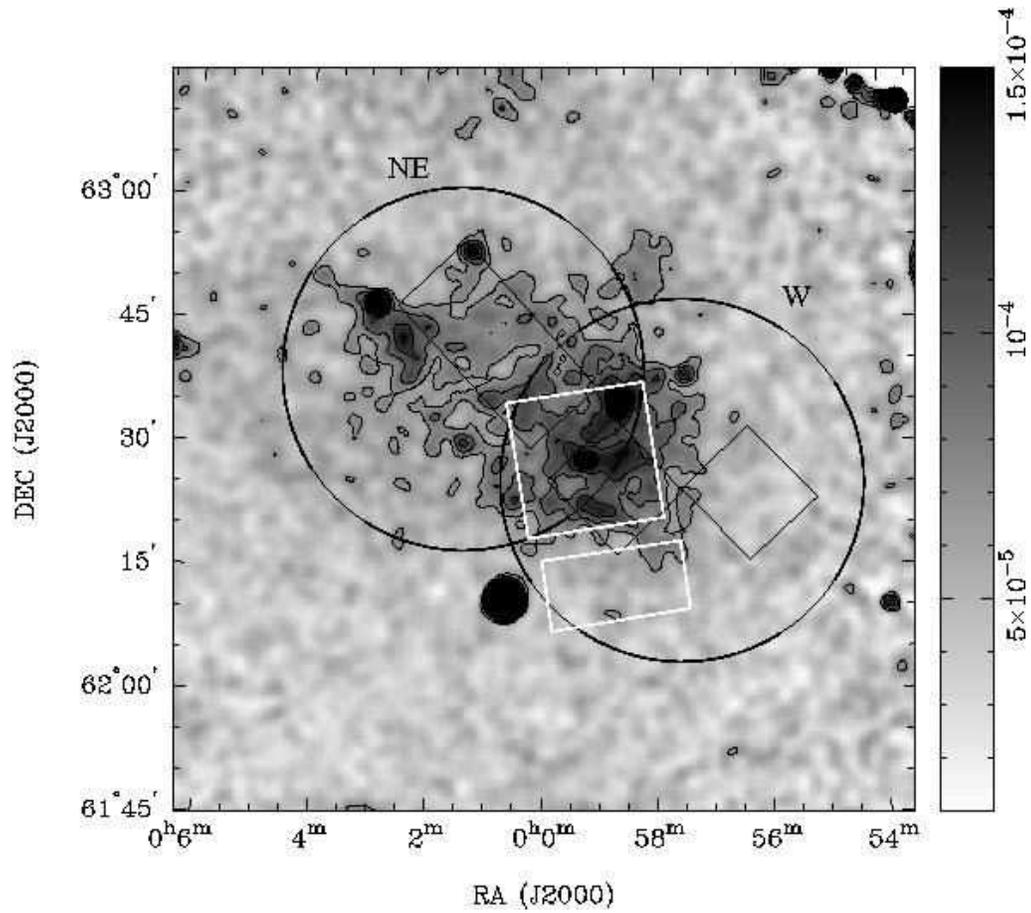}
\caption{Exposure-corrected {\em ROSAT} PSPC  greyscale and contour mosaic image of the
 SNR CTB~1. The image has been smoothed with a 3\arcsec\ FWHM Gaussian and 
units are in counts~arcmin$^{-2}$~s$^{-1}$. The regions covered in two
 separate pointings (NE and W) with the {\em ASCA} GIS
 detectors are marked with circles; black boxes mark the SIS detectors,
 and white boxes mark regions covered with the {\em Chandra} ACIS-I
 (white square) and ACIS-S (white rectangle) detectors.}
\label{fig-image-ctb1}
\end{figure}

\begin{figure}
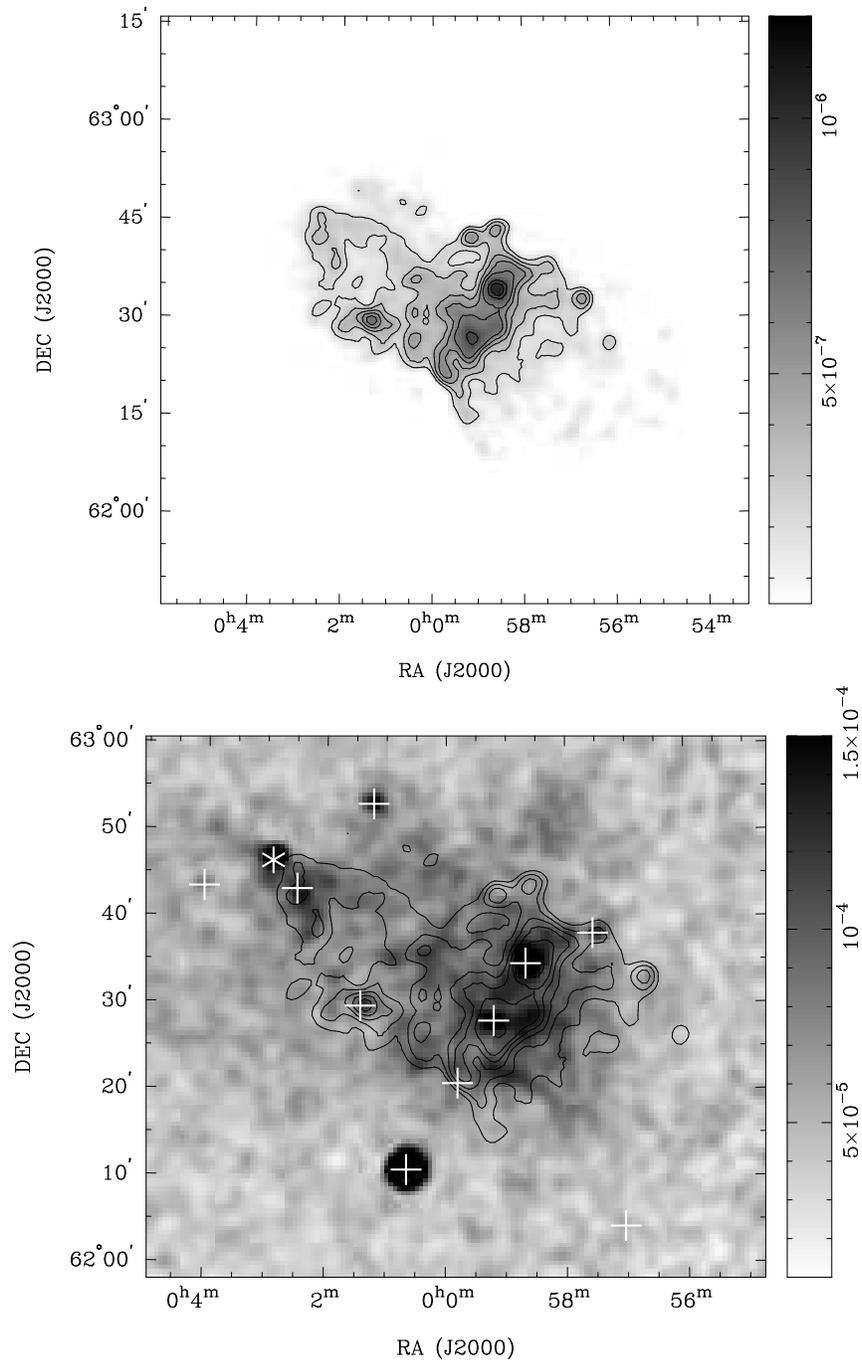

\centering
\includegraphics[height=9cm]{f6a.ps}
\includegraphics[height=9cm]{f6b.ps}
\caption{{\em Top:} Exposure-corrected and background-subtracted {\em
ASCA} GIS mosaic image of CTB~1. The image 
has been smoothed with a 2\arcsec\ FWHM Gaussian and 
units are in counts~arcmin$^{-2}$s$^{-1}$. {\em Bottom:} GIS contours overlaid on the {\em
ROSAT} image from Fig.~\ref{fig-image-ctb1}. Contour levels are:
 20, 30, 40, 50, 60, 80, 90 $\times$ \e{-8} counts~arcmin$^{-2}$~s$^{-1}$. WGA point sources
 are marked with white crosses. The X-ray pulsar RXJ 0002+6246 \citep{hailey95},
located  north-east from the SNR, is marked with a star.}
\label{fig-gis-ctb1}
\end{figure}

In Figure~\ref{fig-image-ctb1} we show the {\em ROSAT} PSPC image
created from the two CTB~1 pointings, as well as regions covered by
 the {\em ASCA} and {\em Chandra} observations. The X-ray 
 emission extends from north-east to
south-west with the brighter emission to the south-west, as shown by
\citet{craig97}. The GIS image and its comparison to  the PSPC image
is shown  in Figure~\ref{fig-gis-ctb1}. There are several point
sources in the field, one of which is an unrelated X-ray pulsar
(undetected in radio band) in the
north-east region outside the SNR radio shell. The two point 
sources  prominent in the center of the SNR
are also obvious in the GIS data.  There is a good overall
correlation between the diffuse SNR emission in the GIS and PSPC
images.   As for HB~21, we produced soft and hard band {\em ASCA} images to investigate possible spectral variations, but found no significant differences. 

\begin{figure}
\centering
\includegraphics[height=12cm]{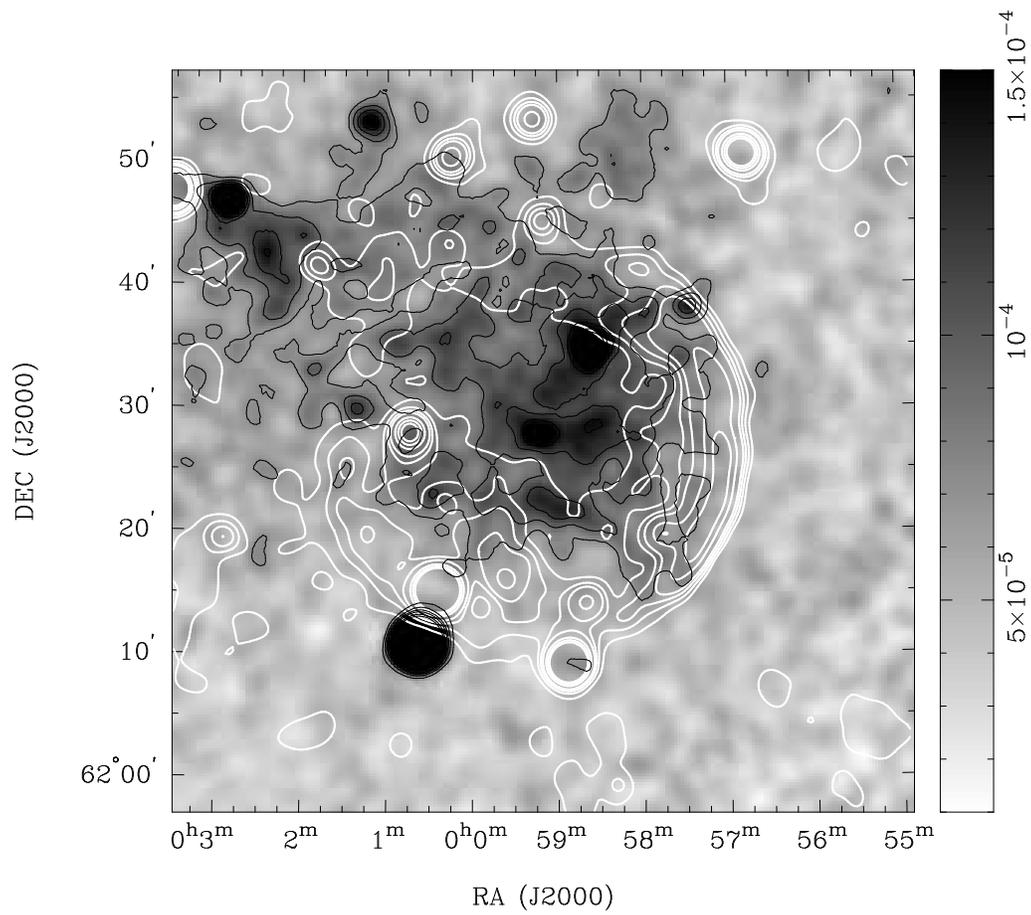}
\caption{{\em ROSAT} PSPC greyscale image of CTB~1 overlaid with the 
92~cm WENSS radio image, which has been 
smoothed to 2\arcmin\ resolution. Contour levels are:
 1, 3, 5, 7, 12, 16, 20 $\times$ 0.27 \jyb .}
\label{fig-radio-ctb1}
\end{figure}

Comparison with the 92~cm WENSS radio image in
Figure~\ref{fig-radio-ctb1} shows that X-ray emission is concentrated inside
the radio shell, and has a break-out morphology in the northeast,
extending further than the radio emission. CTB~1 has almost perfect
spherical radio morphology which  suggests that the blast wave has been
expanding into a relatively uniform  ambient medium, except for in the
north-east. The optical
emission resembles the radio emission, showing  limb-brightened filaments
and diffuse emission in the remnant's interior
\citep[e.g.,][]{fesen97}.  The north-east region has been suggested to be
 a blow-out of  the SNR shock into a lower-density medium, or an unrelated
SNR associated with the X-ray pulsar \citep{hailey95}. The morphology
of optical filaments corresponding to the north-west X-ray emission
supports the blow-out scenario \citep{fesen97}. \hi\ gas in the region
is distributed mostly in the southern half of the remnant, {having an
  arc-like appearance} \citep{landecker82,uyaniker04}, and suggests that the remnant has been influenced by the density distribution of the ambient gas. 

\subsubsection{Spectra}

\begin{figure}
\centering
\includegraphics[height=8cm]{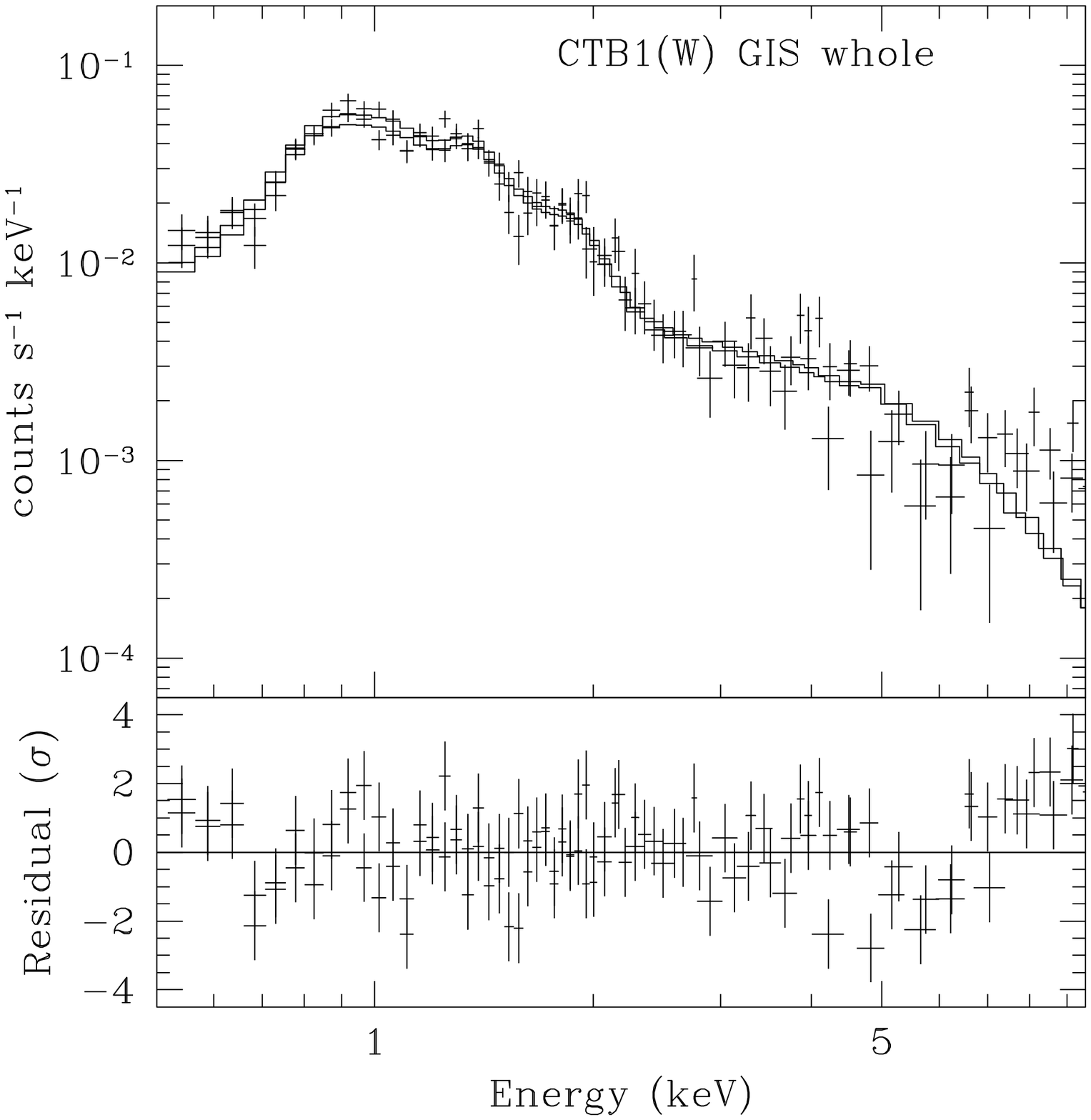}
\includegraphics[height=8cm]{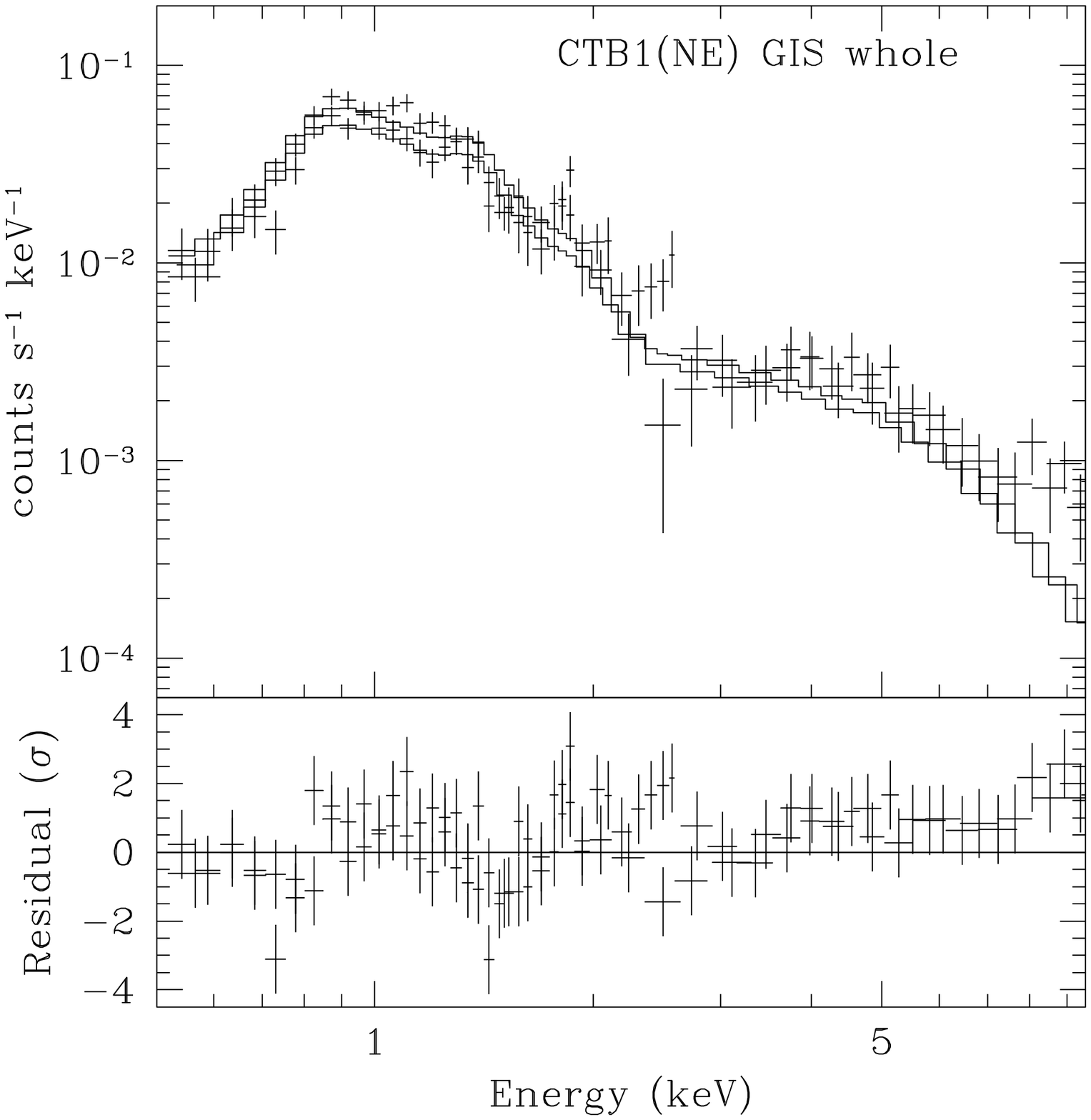}
\includegraphics[height=8cm]{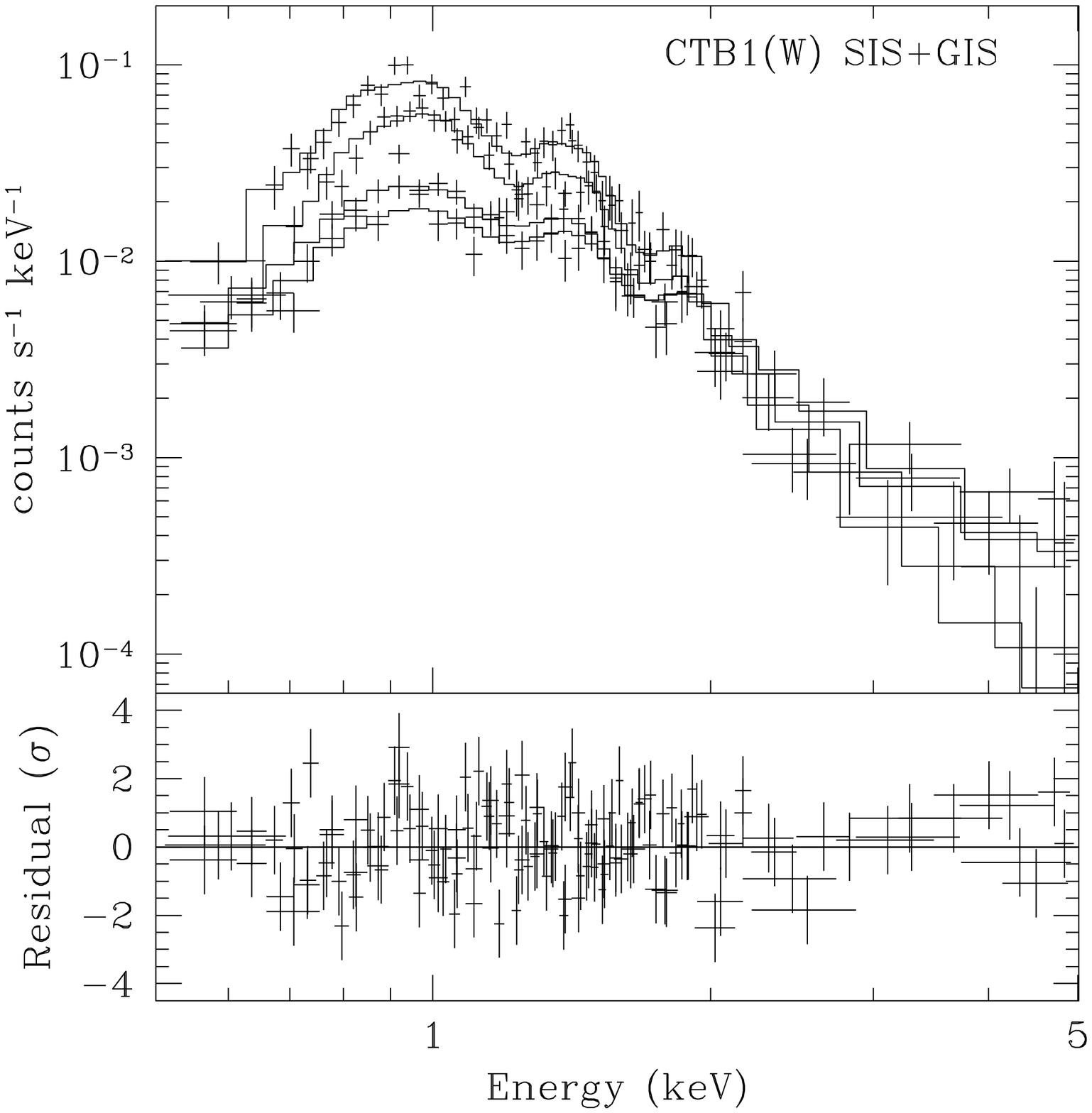}
\includegraphics[height=8cm]{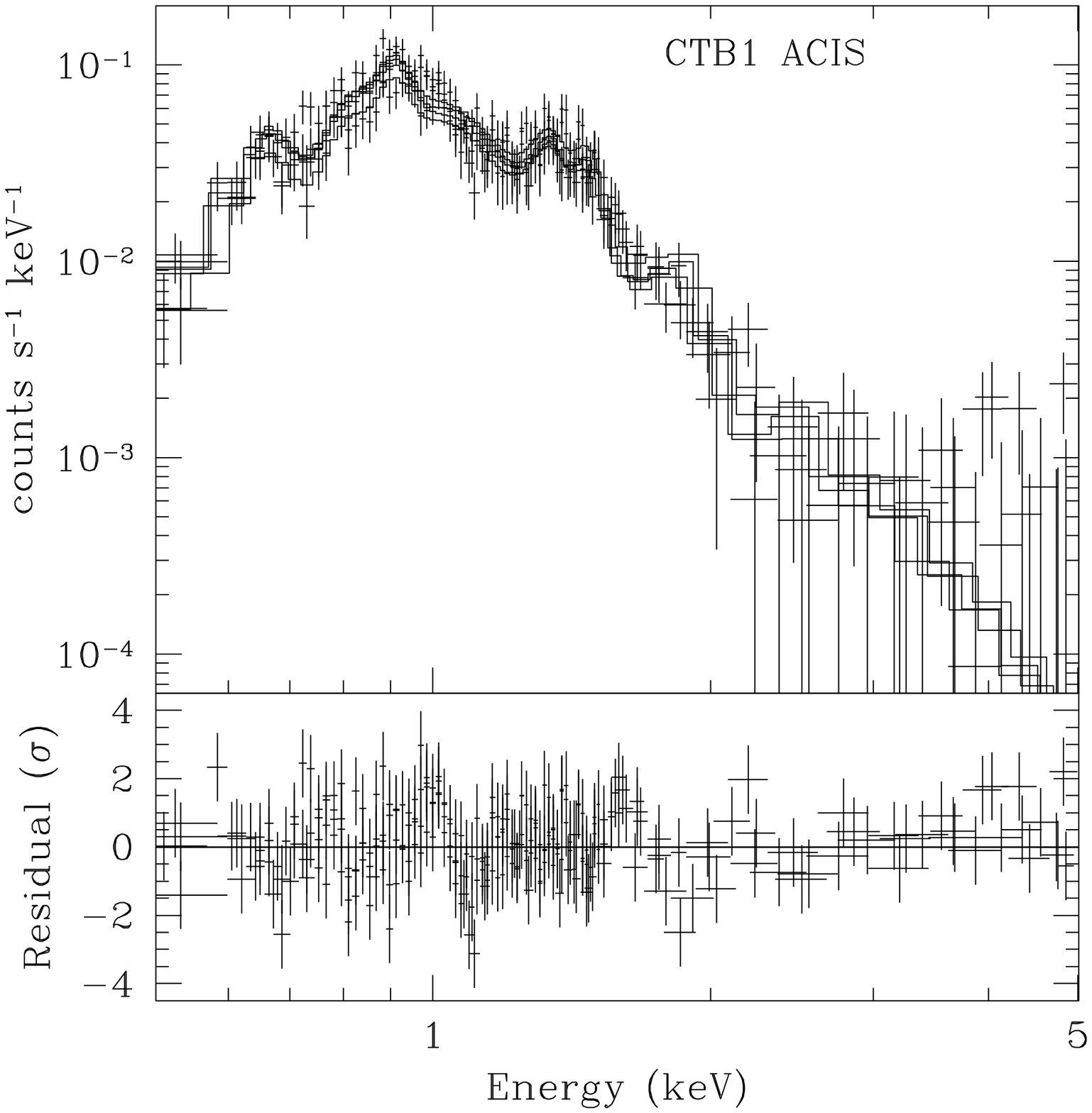}
\caption{{\em ASCA} and {\em Chandra} spectra of the SNR CTB~1 and residuals from the
best-fit models. The GIS and SIS spectra are from the two {\em ASCA}
pointings marked in Figure~\ref{fig-image-ctb1}, and  the ACIS spectra are
 extracted from a 5\arcmin\ square regions located in the inner corner
 of each ACIS chip.  Fit parameters are listed in
Table~\ref{tab-fit-ctb1}. The Mg line is prominent in SIS+GIS spectra of
the W {\em ASCA} region. In addition to Mg, O and Ne+Fe-L lines 
are visible in ACIS spectra. }
\label{fig-spectrum-ctb1}
\end{figure}

The spectra from the GIS regions in the W and NE pointings 
 are shown in Figure~\ref{fig-spectrum-ctb1}. Bright point sources
 have been excluded when extracting the source spectra. No line
 features are prominent in the GIS spectra; the spectra are fitted well
 with a single-component VRAY model with solar abundances. The VPSHOCK
 model gives fit parameters for the NE region consistent with those
 from VRAY model. The SIS and GIS
 spectra from the SIS-covered W pointing are also shown in
 Figure~\ref{fig-spectrum-ctb1}; some indication of a Mg line is
 present. A single-component VRAY model to these spectra resulted in
 large residuals ($\chi_\nu^{2} \approx 2$), and 
addition of a second thermal component improved the
 fit significantly. The two-component model requires a
 cold plasma component with $kT \sim 0.2$~keV, and a warm plasma component
 with $kT \sim 0.8$~keV. The spectra from the SIS regions
 prefer enhanced abundances of Mg, but only in the higher temperature
 component.   The same hard component can be added to the GIS spectra, but does not improve the fit significantly (the improvement is less than $2\sigma$) and it is about 100 times weaker than the soft component. This implies that the hard component is more diluted in the larger area covered by the GIS field of view, and is, therefore, concentrated more in the center of the SNR.  The fit parameters are summarized in Table~\ref{tab-fit-ctb1}.   
The column density and temperature of the cooler plasma component 
we derive here are consistent with those
derived from the previous study with  the {\em ROSAT} data
alone \citet{craig97}.

A {\em Chandra} spectrum has been obtained from four 5\arcmin -square regions from
 four ACIS-I chips, covering the brightest part of the SNR. As
 mentioned before,  blank-sky observations were used for background
 subtraction. We have verified that there is no significant contribution
 from the Galactic ridge emission by comparing the blank sky spectrum
 with that of the local background taken from the source-free region. 
The ACIS spectra are shown in  Figure~\ref{fig-spectrum-ctb1} 
and the spectral fit parameters
 are listed in  Table~\ref{tab-fit-ctb1}. As for W-SIS regions from
  the {\em ASCA} data, the ACIS spectra require a two-component 
thermal model,  but prefer a nonequilibrium VNEI model ($\chi^2=374.4$ for 468 d.o.f.)
 model to VRAY ($\chi^2=433.6$ for 470 d.o.f.). The spectral parameters
 obtained with {\em Chandra}  are consistent with those obtained with
  SIS-covered {\em ASCA} spectra, including the detection of 
enhanced abundances of Mg.
 Lines of O and Ne+Fe-L blend are also clearly visible in the ACIS spectra.
 As in the case of HB~21, detection of enhanced abundances in CTB~1
 implies the presence of ejecta in this SNR. 

There are several point sources from the WGA {\em ROSAT} Point Source 
catalogue \citep{white00} towards CTB~1. One of the sources is the unrelated X-ray pulsar
 RXJ 0002+6246  \citep[1WGA J0002.9+6246][]{hailey95} located
outside the SNR shell in the north-eastern region.  We detected $120\pm40$
 background-subtracted counts within the 6\arcmin-radius circle at the location of 
the X-ray pulsar, but no compact source is obvious in the GIS data. 
Other WGA sources also have a low number
of counts, and we are unable to investigate their nature in more detail.


\subsection{HB 3 (G132.7+1.3)}

\subsubsection{Morphology}

The {\em ROSAT} PSPC image, composed of four pointings towards HB~3, is
shown in Figure~\ref{fig-image-hb3}, including the SNR regions covered
by the {\em ASCA} and {\em XMM} detectors. The PSPC image has been 
smoothed with a Gaussian of 5\arcsec\ FWHM. The X-ray emission is 
brightest in the center,
with fainter emission extending to the north and south. 
As noted by \citet{leahy85}, the  central
emission appears as a ring with a radius of  $\sim 16\arcmin$.
 The GIS image and its overlay with the {\em ROSAT} image are shown 
in Figure~\ref{fig-gis-hb3}.   The GIS image also shows bright 
 emission corresponding to the PSPC ring.  The 
X-ray emission from the north region shows good
correlation between  GIS and PSPC data, and there is a prominent bright clump
between the central and north pointing.  As for HB~21 and CTB~1, we produced soft and hard band {\em ASCA} images to investigate possible spectral variations, but we found no significant differences.  In the south region, however,
 the {\em ASCA} observations reveal the presence of a compact 
source, possibly coinciding with the {\em ROSAT} All Sky Survey source 
1RXS~J021703.6+620136 (within the position error of 14\arcsec). We analyze 
this source in more detail in section \S\ref{sec-ps_hb3}.

\begin{figure}
\centering
\includegraphics[height=14cm]{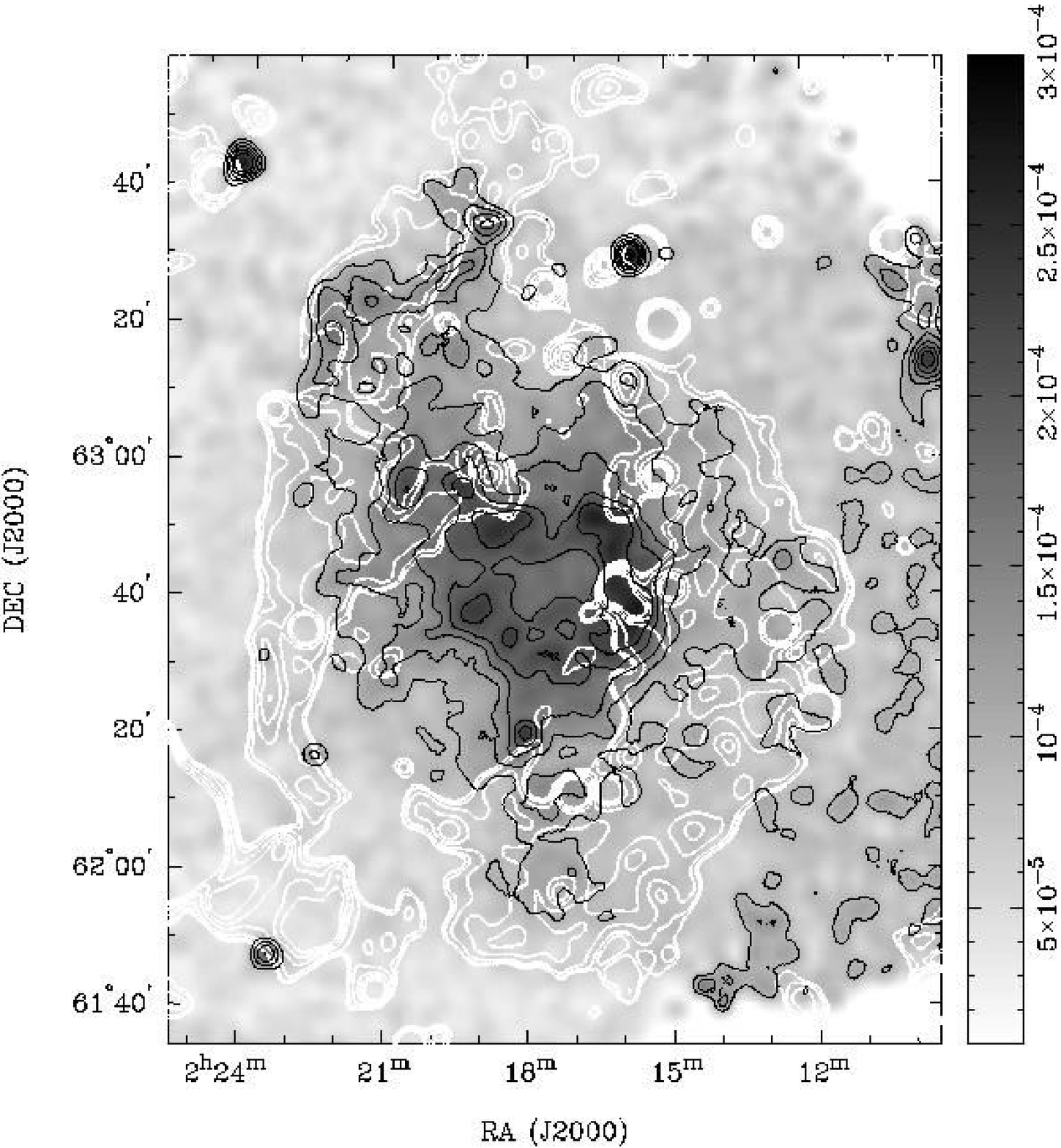}
\caption{{\em ROSAT} PSPC greyscale image of HB~3 overlaid with the 
92~cm WENSS radio image, which has been 
smoothed to 3\arcmin\ resolution. Contour levels are:
 3, 10, 30, 50 $\times$ 0.01 \jyb .}
\label{fig-radio-hb3}
\end{figure}

The comparison between the radio and X-ray emission is shown in
 Figure~\ref{fig-radio-hb3}.  As mentioned before, HB~3 is located
 close to the W3 \hii\ region, which can be partially  seen  in the
 south-east corner of the radio image in Figure~\ref{fig-radio-hb3}.  There is not
 much correlation between the radio and X-ray morphologies, as noted before
 by \citet{leahy85}.  Diffuse X-ray emission extends to the north radio
 shell, and weaker diffuse emission extends to the southern radio
 limb.  It is not clear whether this southern emission belongs to
 the SNR, or to a newly detected compact {\em ASCA} source.
\citet{routledge91} suggested that the fainter diffuse
 emission at the SNR's north part is caused by a density gradient
 produced by evolution in a cavity, rather than by a Galactic plane
 gradient, because the elongation of the SNR is not perpendicular to
 the Galactic plane ($\sim 75$\degr). The bright inner SNR ring was
 suggested to be  due to a second SNR, a pre-supernova circumstellar
 shell, or an  unusual density distribution in the ejecta \citep{leahy85}. 
  The lack of X-ray emission in the
 south-eastern  region was attributed to absorption along the line of
 sight to the remnant \citep{leahy85}, which  was supported 
by the detection of a CO cloud \citep{huang86}, suggested to be
 interacting with the remnant on the basis of morphological agreement
 \citep{routledge91}. \oh\ masers, which are recognised as the
 signpost for molecular cloud interaction with an SNR
 \citep[e.g.,][]{wardle02}, have been detected towards the W3/HB~3
 complex. One of the masers is located very close to the SNR, with a velocity
 close to the systematic velocity of the gas associated with the remnant;
 the maser is, however, located outside the SNR shell and is believed
 to be associated with star-forming phenomena rather than the SNR shock
 \citep{koralesky98}.  Optical emission from the SNR was found to be well-correlated with
 the radio emission, with a multiple shock structure found in the
 western SNR shell and lack of emission in the southeast region \citep{fesen95}.

\subsubsection{Spectra}
\label{sec-xmm-spec}

\begin{figure}
\centering
\includegraphics[height=12cm]{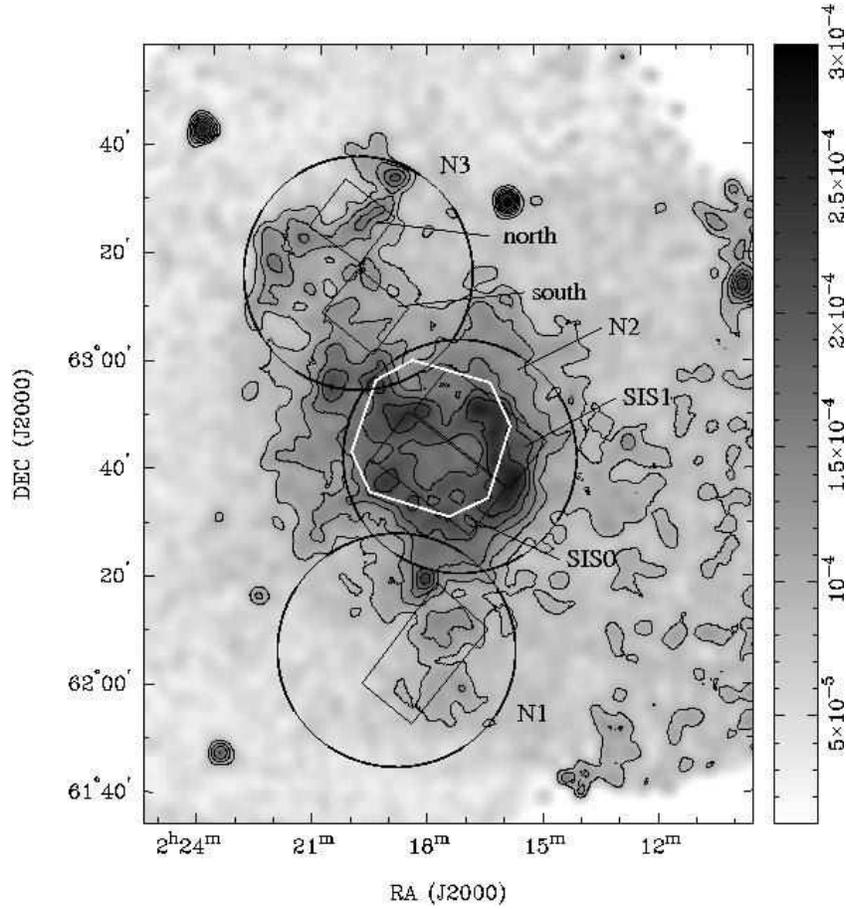}
\caption{Exposure-corrected {\em ROSAT} PSPC  greyscale and contour mosaic image of the
 SNR HB~3. The image has been smoothed with a 5\arcsec\ FWHM Gaussian and 
units are in counts~arcmin$^{-2}$~s$^{-1}$. The regions covered in three
 separate pointings (N1, N2 and N3) with {\em ASCA} GIS
 detectors are marked with circles, black boxes mark SIS detectors,
 and a white box marks region covered with {\em XMM-Newton} pn detector.}
\label{fig-image-hb3}
\end{figure}

\begin{figure}
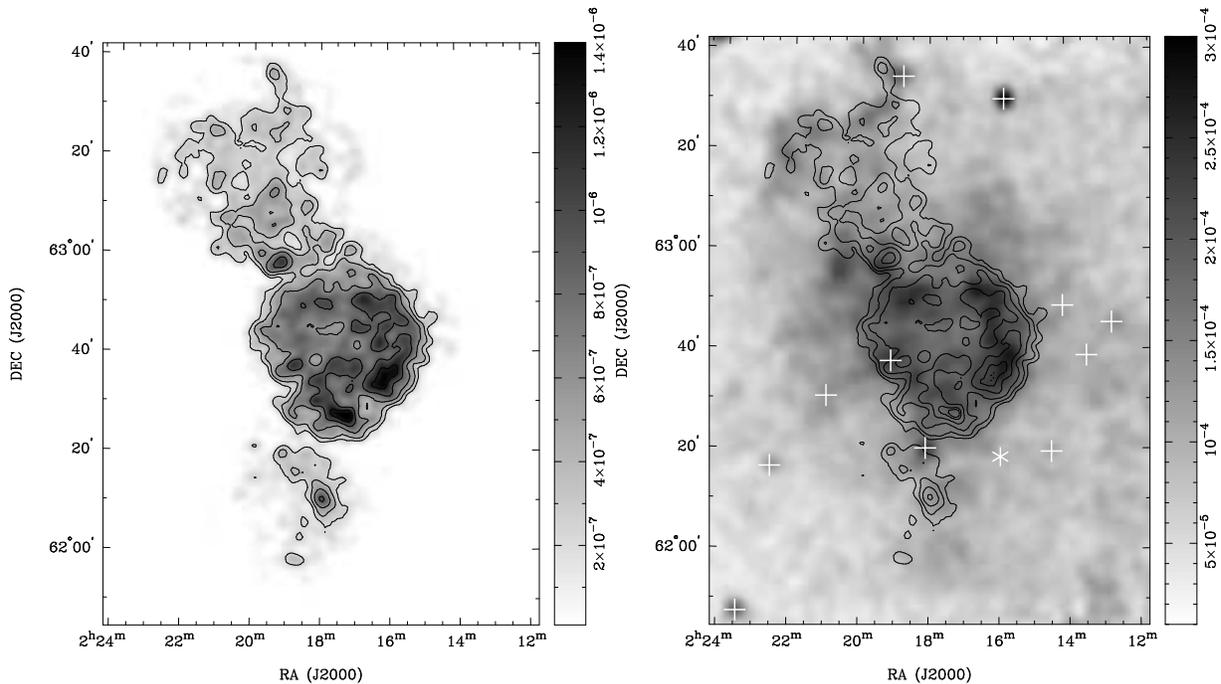

\centering
\includegraphics[height=9cm]{f10a.ps}
\includegraphics[height=9cm]{f10b.ps}
\caption{{\em Left:} Exposure-corrected and background-subtracted {\em
ASCA} GIS mosaic image of three pointings towards HB~3. The image has
been smoothed with a 2\arcsec\ FWHM Gaussian and 
units are in counts~arcmin$^{-2}$~s$^{-1}$. {\em Right:} GIS contours overlaid on the {\em
ROSAT} image from Fig.~\ref{fig-image-hb3}. Contour levels are:
 15, 25, 40, 50, 60, 80 $\times$ 1.5\ee{-8}
counts~arcmin$^{-2}$~s$^{-1}$. The WGA point sources
 are marked with white crosses. The unrelated radio pulsar J0215+6218
\citep{lorimer98}, located in the south-western SNR region, is marked with a star.}
\label{fig-gis-hb3}
\end{figure}

\begin{figure}
\centering
\includegraphics[height=8cm]{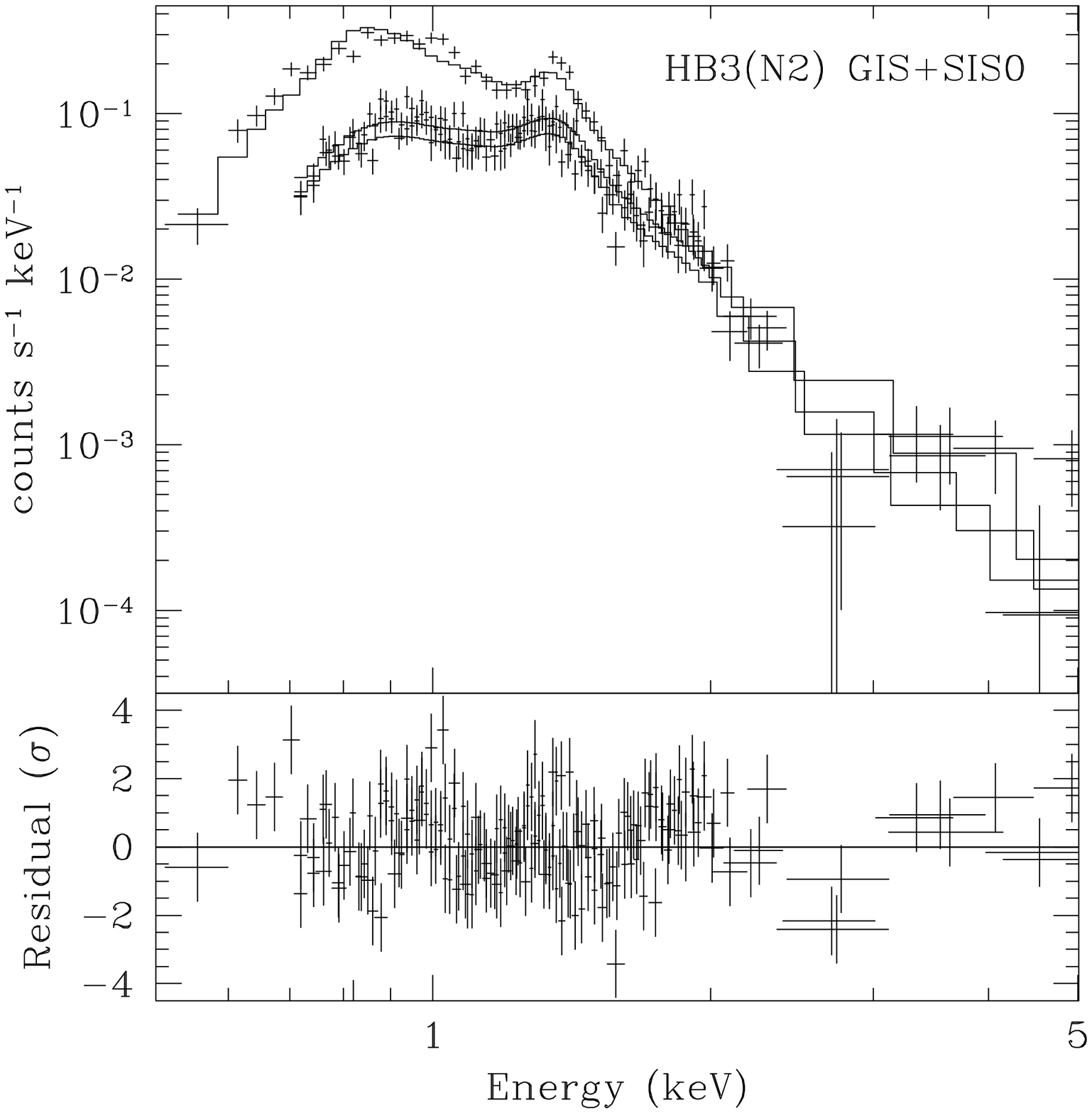}
\includegraphics[height=8cm]{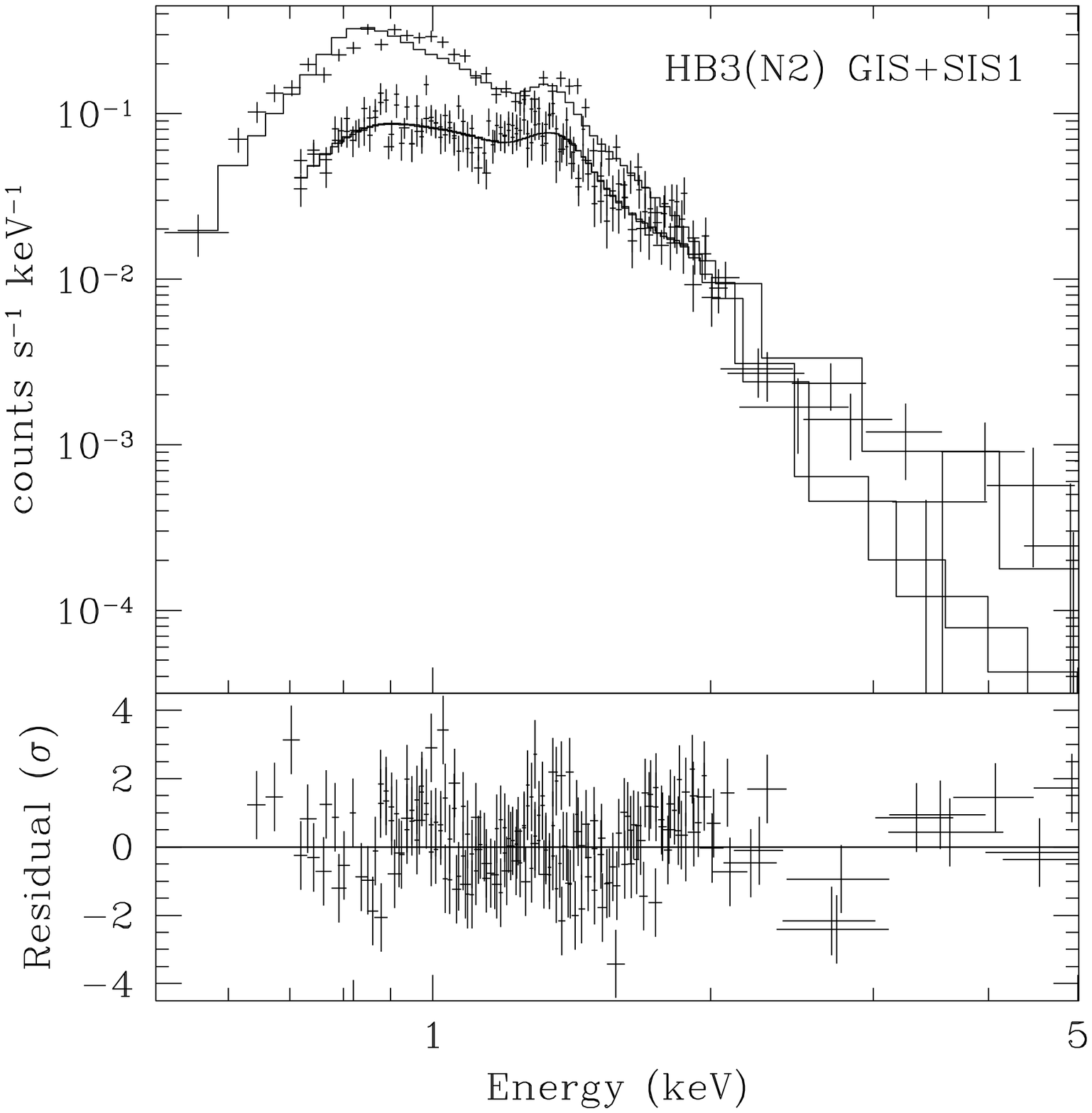}
\includegraphics[height=8cm]{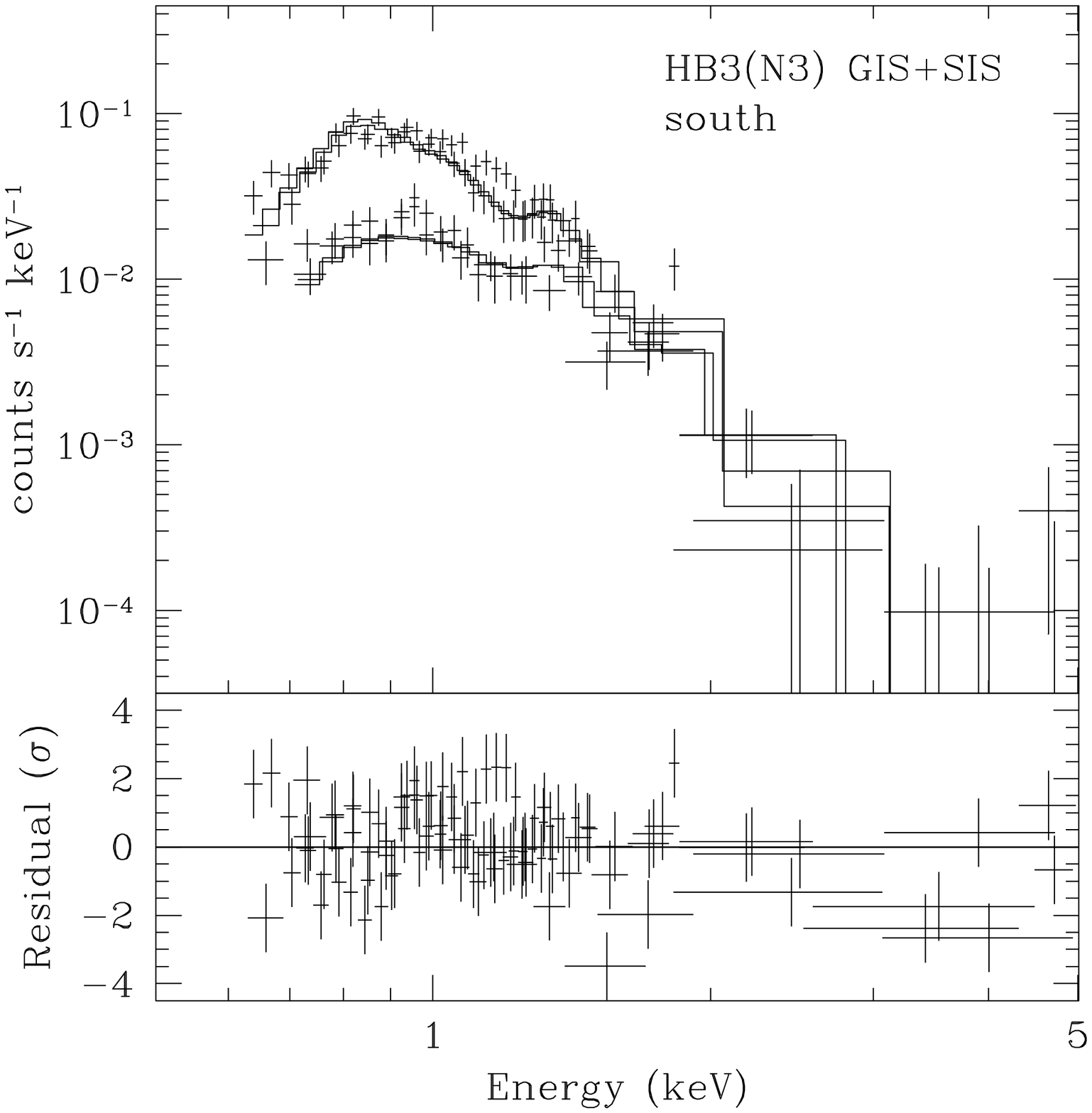}
\includegraphics[height=8cm]{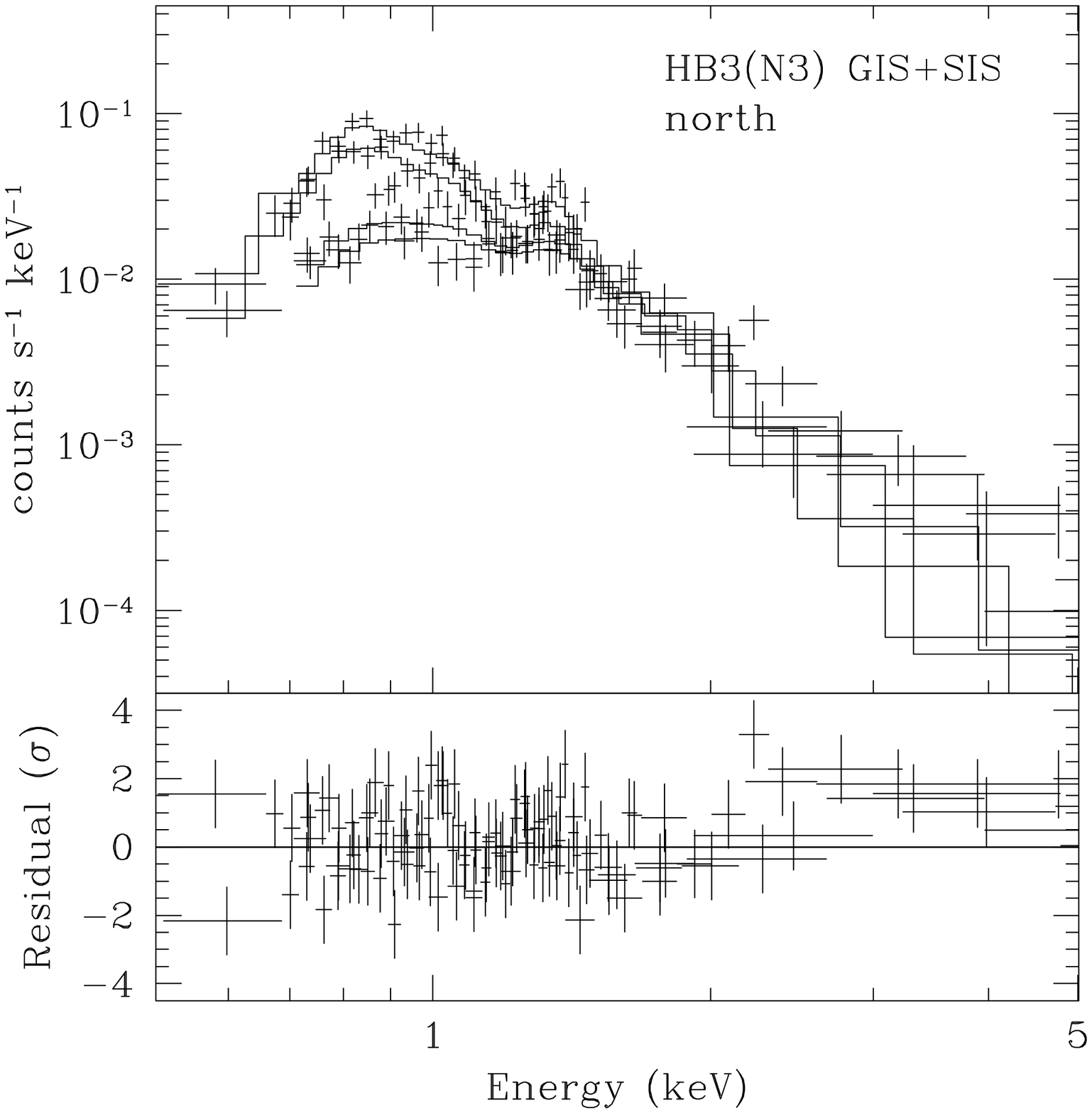}
\caption{GIS and SIS spectra from the two {\em ASCA}
pointings towards HB~3 marked in Figure~\ref{fig-image-hb3}, and residuals from the
best-fit models. Fits parameters are listed in
Table~\ref{tab-fit-hb3}. Mg line is prominent in the spectra from N2 pointing.}
\label{fig-spectrum-hb3}
\end{figure}

The GIS and SIS spectra from the two 
SIS regions (SIS0  and SIS1 cover different regions; see
Fig.~\ref{fig-image-hb3}) in the central pointing towards HB~3 are shown in
Figure~\ref{fig-spectrum-hb3}. Bright point sources
 have been excluded when extracting the source spectra. 
 The VRAY model with enhanced Mg  abundances
was able to give a satisfactory fit to the spectra from the central
 SIS0 pointing, but the region corresponding to the SIS1 pointing does not
require enhanced abundances. For completeness, we list the Mg
abundances for both SIS pointings in
Table~\ref{tab-fit-hb3}, together with other fit parameters.   
The joint fit to the GIS and SIS spectra from the north
pointing was performed for the two SIS CCDs separately 
(marked N3-south and N3-north in Fig.~\ref{fig-image-hb3}). 
We also applied the VPSHOCK and VNEI models to the N3 regions, and the
spectral parameters were consistent with those from the VRAY model.
The spectral fits to the N3-north region give solar 
abundance and  higher plasma temperature than the central SNR
region.  That is also the case for the N3-south region, 
with a difference that the column density is slightly 
lower than in the central region. If we freeze $N_{\rm H}$ to the value
from the N2 regions when fitting the N3 regions, we get a plasma temperature
 consistent with the N2 pointings for N3-north, while N3-south still
 requires a slightly higher temperature.  
\citet{leahy85} reported that the plasma temperature decreases  
from 1~keV in the center, to 0.3~keV towards the edge. We find a
 low temperature in the SNR center with the {\em ASCA} data 
 and that the plasma temperature, and possible the column density, varies
 non-uniformly  from the center of the remnant to the north.

\begin{figure}
\centering
\includegraphics[height=12cm]{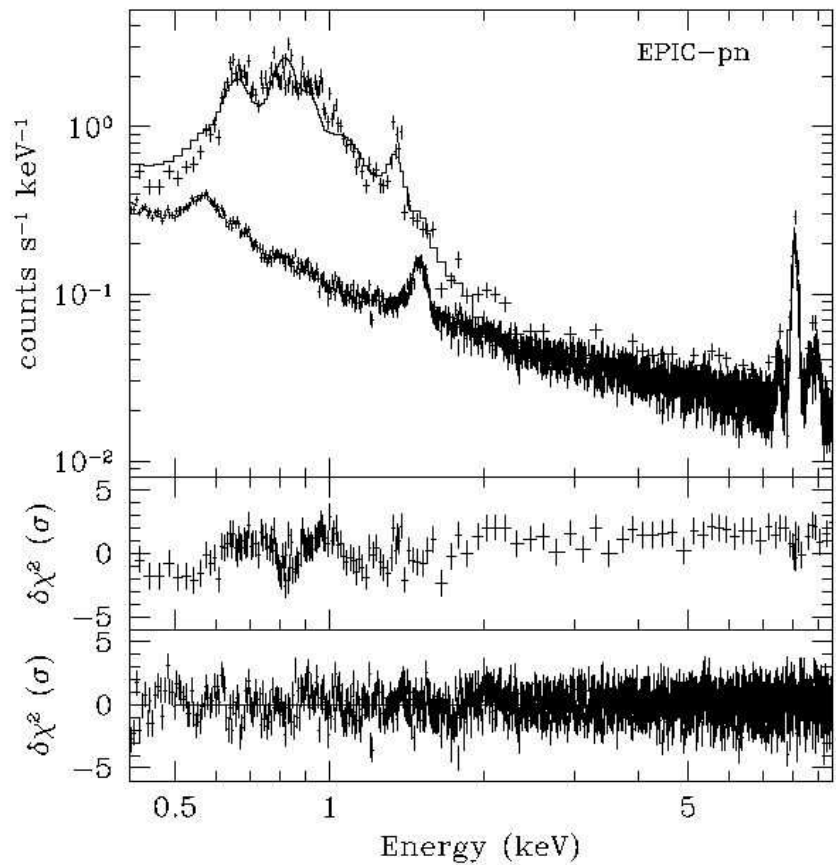}
\caption{{\em XMM-Newton} EPIC-pn spectrum from the 5\arcmin -radius central region of
HB~3. The spectrum of the source is fitted simultaneously with the
background spectrum: the solid line in the upper panel represents source+background spectrum
and its residuals from the best-fit model are given in the middle
panel; the dashed line in the upper panel represents
background spectrum and its residuals are given in the lower panel. The strong lines
visible in the background spectrum are Al-K line at 1.5~keV and
Cu-Ni-Zn-k line complex around 8~keV. }
\label{fig-xmm-hb3}
\end{figure}

An {\em XMM} spectrum obtained from the 5\arcmin -radius region from the
 central region of the EPIC-pn detector is shown in Figure~\ref{fig-xmm-hb3}. No
 spectra were obtained from the EPIC-MOS detectors due to a very low number
 of counts in these two detectors.  Because the diffuse emission from
 the remnant covers the whole detector, we used the blank sky
 observations for background subtraction. The source and background
 spectrum were fitted simultaneously. The background was modeled with
 two thermal components using the VRAY model with
 temperatures 0.08 and 0.2~keV \citep{lumb02}, in combination with a
 power low model and zero-width Gaussians to account for the detector
 background. The SNR emission was modeled with the VRAY or VNEI model, which was
 not applied to the background spectrum. We achieved a good fit
  with a single-temperature plasma component with enhanced Mg
 abundance ($\chi_\nu^{2}$=1.21), and the fit parameters are
 consistent with those from the central {\em ASCA} (N2) pointing. 
 However, some residuals were still present in the {\em XMM} spectrum
 in the 0.6--1.0~keV range. This is the energy range where O and Ne lines
 are present, as well as Fe L-shell lines \citep[e.g.,][]{masai97,brickhouse00}. 
 Therefore, we thawed the O elemental abundance next, in addition to Mg, which  improves the
 fit significantly.  We list parameters of this fit in
 Table~\ref{tab-fit-hb3}. However, if the Ne abundance is thawed next,
 the fit is also improved significantly, but the column density and
 temperature change considerably (\nh~$\sim 0.5$\cm{-2} and $kT \sim
 0.3$~keV) and the fit values are not consistent with {\em ASCA} results
 anymore. Thawing Fe abundances no longer improves the fit. 
On the other hand, if Fe is thawed first, in addition to 
Mg alone, the fit is also improved and
  results in slightly enhanced Mg and under-abundant Fe. However,
  the column density and temperature also change to above values.
Therefore, we can not determine with certainty the elemental
 abundances of Ne and Fe, as well as the \nh\ and $kT$ values sampled by
 the {\em XMM} region. These issues will possibly be resolved with
 new data obtained recently with {\em XMM} towards four regions of
 HB~3, including the central region.

In summary, the {\em ASCA} data for HB~3 indicate that there is an
 enhanced abundance of Mg present only in the center of the remnant.
  The {\em XMM} data suggest a few possibilities, including enhanced
 abundances of Mg, Ne and O, or enhanced abundances of Mg and
 sub-solar abundances of Fe. It is not too
  surprising to have different fit parameters 
from the  {\em XMM} and {\em ASCA}
 spectra, because {\em XMM} is more
 sensitive than {\em ASCA} at lower energies, and the regions sampled
 by the two observatories differ somewhat.  As for the origin of the inner 
ring in HB~3, our data are not sensitive enough to discriminate between the 
scenarios suggested above.

\subsubsection{Point sources in HB~3 field}
\label{sec-ps_hb3}

There are a few point sources identified from the WGA catalogue present
in the field towards HB~3, but none of them have enough counts in
the {\em ASCA} data to be analyzed here. No X-ray emission was
detected from the radio pulsar PSR J021800.0+621000, which is not
surprising considering its age \citep[$\sim$\e{6} yr;][]{lorimer98}. 

\begin{figure}
\centering
\includegraphics[height=8.5cm]{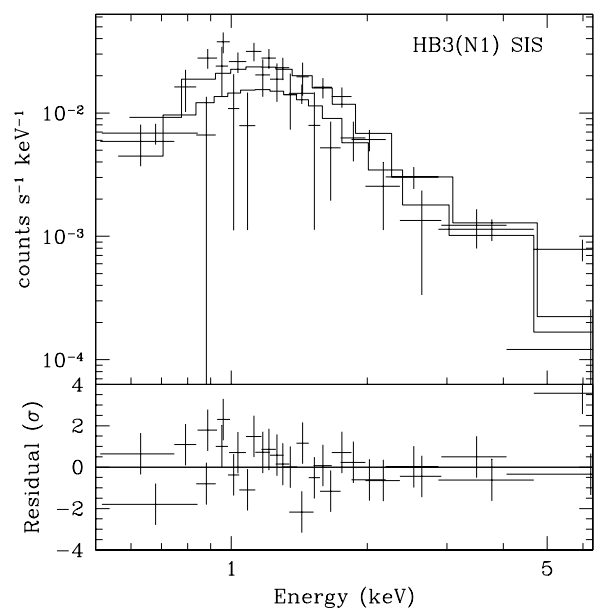}
\caption{SIS spectra from the compact source located in the south N1
pointing towards HB~3, and residuals from the
best-fit power law model. Fits parameters are listed in section \S ~\ref{sec-ps_hb3}.}
\label{fig-spectrum-hb3-n1}
\end{figure}

To investigate the nature of the bright southern source detected with
{\em ASCA} we used only the SIS spectra, shown in 
Figure~\ref{fig-spectrum-hb3-n1}, because the SIS detectors have a
smaller point spread function than the GIS detectors. The source spectra were
extracted from the 4\arcmin\ circular region around the compact
source, resulting in about 200 background subtracted counts, and the background 
spectra are extracted from the thin
ring-like region around the source region. While a thermal model of
plasma in equilibrium (RAY), and also a bremsstrahlung model 
 can be rejected because of  statistically  unacceptable fits
 ($\chi_\nu^{2} \sim 3$), both a power law and blackbody models give reasonably
 good fits to the source spectra. The fit parameters for the power law
 model are: a column density 
 of $N_{\rm H}$=2.9 (1.8--4.1)\ee{21}\cm{-2}, photon index 
 of $\Gamma$=2.74 (2.39--3.16), and $\chi_\nu^{2} =1.60$ 
for 27 degrees of freedom.  The fit parameters for the blackbody
 model are: $N_{\rm H}$=1.1 (0.4--2.1)\ee{21}\cm{-2}, temperature 
 of $kT$=1.91 (1.35--2.75)~keV, and $\chi_\nu^{2} =1.90$ for 27 degrees of freedom. An 
unabsorbed X-ray flux in the 0.5--10.0~keV band is, for both models,
\e{-12}\ergs. The column density appears to be, for both cases,  
 lower than from the SNR, suggesting that the source is probably not
 at the same distance as the SNR.

The photon index from the compact source is softer than 
that of classical young pulsars which are
 found with photon index values between 1.1--1.7; it is also harder than the
 photon index values found in compact central objects (CCOs) 
and anomalous X-ray pulsars (AXPs) with a photon index of $\sim 4$ \citep[e.g.][]{chakrabarty01}.
There is a small-diameter source  present in the WENSS radio
  image that coincides with the location of the {\em ASCA}
source. \citet{landecker87} derived a steep radio spectral index of
$-0.98\pm0.8$, using observations at 408~MHz and other available
frequencies.
The field towards this region is crowded with optical and
  infrared source, but since we do not know the exact position of 
the possible point source,  we cannot determine if there is an 
optical or infrared counterpart to the {\em ASCA} source.
Further observations are needed to
establish the exact nature of this source.


\section{Discussion}

\subsection{SNR properties}

Our study confirms the thermal nature of the central X-ray emission
in three SNRs ---  HB~21, CTB~1 and HB~3 --- classified as mixed-morphology
SNRs. In addition, we detected enhanced abundances in all three SNRs.

HB~21 shows a flat temperature  of around 0.6~keV across its center, 
which is similar to value often found in MM SNRs \citep{rho98}. 
Unlike for the other two SNRs studied here, there are no {\em ASCA} 
pointings towards the edge of the SNR shell in HB~21, and therefore, 
 we do not have a complete knowledge of the temperature distribution
 in this remnant. {\em ASCA} data  show a  possible column density
 variation in HB~21, with higher column density in the P2 pointing. This 
is consistent with a suggested interaction with molecular clouds
 in that region. There are bright radio indentations north and south 
from the P2 pointings and this is where \citet{koo01} detected broad
 molecular lines. Interactions with dense material may
   have resulted in the formation of a radiative shock in these regions, 
which would explain why there is no X-ray
emission corresponding to these bright radio regions, while there is
X-ray emission corresponding to the radio shell extending to the west, 
right next to it. The spectrum of HB~21 is dominated by the Si line, and
we found enhanced abundances of Si and S in the center of the SNR.

Observations of CTB~1 reveal two plasma components in this remnant, 
one around 0.2~keV with solar abundances, and the other around 
0.8~keV with enhanced Mg. Line features between 0.6 and 1.5~keV are visible in the 
{\em Chandra} spectrum of CTB~1. We found enhanced abundances of
Mg for the higher temperature component only from SIS and ACIS 
regions in CTB~1, implying the presence of ejecta in the center of
CTB~1. The presence of two thermal components is similar to what is seen in
{\em ASCA} observations of IC~443 \citep{kawasaki02}, although without
enhanced abundances. The higher sensitivity of the IC~443 data allowed
spatial identification of its spectral component as a 
cool X-ray shell and warmer central region. 
 Indeed, IC~443 and CTB~1 share similar morphology, both having a
 blow-out region on one side and interaction with dense \hi\ material
 on the other side. The 0.2-keV component in CTB~1 could thus originate from 
 a cooled SNR shell, since we see X-ray emission in CTB~1 extending up
 to the radio shell, mostly in the north-west side. \hi\ material is
  densest in the south-east region and the radio
 continuum is brightest here \citep{landecker82}, so it could be that
  the SNR shock cooled down significantly here due to an encounter with
   the denser material. On the other hand, the plasma in CTB~1 might
     be similar to that in W28, where two distinct plasma  components 
are found 
  in the center of the SNR, although without enhanced
  abundances; the soft plasma component here was suggested to
  originate from denser clumps embedded in diffuse hot gas \citep{rho04}.

For HB~3 the situation is not so clear. {\em ASCA} observations
indicate a small but significant column density and 
plasma temperature variation from the center of the SNR towards the northern shell.
There seems to be a single plasma component present, with 
enhanced elemental abundances of Mg in the center of the SNR. 
  However {\em XMM} data allow for a different possibility,
  which results in a lower column density in the center and a higher
  plasma temperature than derived from the {\em ASCA} data, 
as well as additional enhanced abundances of O
  and Ne or sub-solar abundances of Fe.

\subsection{MM models}

A characteristic feature of MM SNRs is a flat 
temperature profile. Due to the low
number of counts in the SNRs studied here, we are unable to derive the
radial temperature distributions. Nevertheless, we
consider different models  that can produce centrally-concentrated
X-ray emission in the three SNRs studied here.
 
\subsubsection{Evaporating clouds}
\label{sec-evap}

One of the earliest MM SNR models invokes evaporation of clouds which
are left relatively  intact after the passage of the SNR blast wave
\citep[e.g.,][]{white91}.  In such a case most of the X-ray emitting
gas is due to dense embedded molecular material that  evaporates
in the SNR interior due to saturated thermal conduction in the hot
gas.  This model depends on the $C / \tau$ ratio, where $C$ is the ratio of
the mass in the clumps to the mass of intercloud medium, and $\tau$ is
the ratio of the cloud evaporation time to the SNR's
age. Ratio $C / \tau = 3$--5 is found to be the most appropriate
for the steep radial surface brightness profiles of MM SNRs \citep{white91}.  The remnant
is assumed to still be in the adiabatic phase, and the post-shock
temperature is thus related to  the observed temperature of the X-ray gas
in the Sedov phase,  scaled by a $K/q$ factor which depends on the $C / \tau$ ratio
\citep{white91}: 
\begin{equation}
T_s = T_{\rm Sedov} \times (K/q) = 0.78~kT_x~(K/q). 
\end{equation}
 Typical $K/q$ values for $C / \tau$=3--5 are 1.0--1.3. 
Note that $T_s$ refers to the case where electron-ion equilibrium
  is assumed.
The Rankin-Hugoniot relation for the temperature of shocked gas 
with adiabatic index $\gamma=5/3$ 
is used to derive the velocity of the blast wave:
\begin{equation}
v_{s} = \left ( \frac{16 k T_{s}}{3 \mu m_H} \right ) ^{1/2},
\end{equation}
where $\mu=0.604$ is the mean atomic weight, and the age of the
remnant is:
\begin{equation}
t=2 r_{s} / 5 v_{s}.
\end{equation}
The explosion energy is found from
\begin{equation}  
E = \frac{E_{\rm Sedov}}{(\gamma + 1)K} 
  = \frac{16 \pi (1.4 n_0 m_{\rm H})}{25(\gamma + 1)K} \frac{r^{5}_{s}}{t^2}.
\end{equation}
The SNR dynamical parameters  for the central SNR pointings obtained from this model are listed in
Table~\ref{tab-results}. Although significant evidence exists that CTB~1 and HB~3 are in the radiative phase of evolution and, thus, the evaporating model is unlikely scenario for them, we list parameters of this model for all the SNRs for completeness. The SNR parameters used for calculations, 
 like shock radius $R$, emission volume $V$, distances $d$ to the SNRs 
are also listed in Table~\ref{tab-results}. The post-shock electron 
densities $n_{\rm H}$ are derived from X-ray emission measure 
$EM = n_{\rm H}^{2} f V/ 4 \pi d^2$.  The emission volume $V$ was taken to
be a slab whose area corresponds to the spectral extraction region, 
and for depth we took half of SNR diameter, the later being a simple 
way to account for the fact that X-ray emission does not 
fill the whole remnant.   For comparison, we also list the values
obtained using Sedov solutions in Table~\ref{tab-results}. 
The observed densities for Sedov solution range 
 from low values between $\sim 0.07$\cm{-3} in HB~3 to more typical values of
 $\sim 0.4$\cm{-3} in CTB~1. We note that
  these values may not represent the immediate post-shock densities for
   the Sedov solution, but we carry out the calculations to illustrate how SNR
   parameters change depending on the SNR model applied. The SNR age 
ranges from 5\ee{3} years in HB~21 to 
 3\ee{4} years in HB~3. The explosion energy derived
from Sedov solutions is more than one order of magnitude 
lower  from the canonical explosion energy of \e{51} ergs. The explosion energy
derived from the evaporating cloud model is higher than for the Sedov
model, but still an order of magnitude lower than \e{51} ergs in most cases.  
 This is often the case when evaporation model is applied to 
 MM SNRs \citep[e.g.,][]{harrus97,slane02}.


\subsubsection{Radiative SNRs}
\label{sec-rad}

Another model has been applied to explain the emission for the mixed
morphology SNR W44 \citep{cox99,shelton99} as due
to the effect of thermal conduction in the interior of the radiative 
remnant that is expanding into a density gradient.  In the
radiative phase of SNR evolution the shock temperature falls below
$\sim$0.1\,keV, and there is no X-ray emission above $\sim$1\,keV
from the SNR shell. The soft X-ray emission from the radiative shock
is readily absorbed, and so the central hot region of the SNR
dominates the X-ray emission. The Sedov model for this phase gives a
high temperature but low density in the center of the remnant which
would result in low emission from the center. Thermal conduction via Coulomb
collisions between the electrons and ions inside the hot plasma
 \citep{spitzer56} can reduce the central temperature and increase the
 central density, which can dramatically increase the brightness of
the central emission. Thermal conduction was thought not to be
significant in SNRs because of the presence of strong magnetic fields,
but it is possible that the magnetic fields get swept up from the SNR interior which
enables efficient thermal conduction.

We examine the possibility that the three SNRs studied here 
are in the radiative
phase. All three SNRs are found with dense gas present in their
vicinity, as suggested by \hi\ and CO observations. 
The radiative shell size $R_s$ and shell formation time $t_{\rm
  shell}$ for an SNR can be found from \citep{cox99}:
\begin{equation}
R_s \approx 12 E_{51}^{1/8} t_{\rm sh,20}^{3/4} {\rm\ pc},
\end{equation} 
\begin{equation}
t_{\rm shell} = 53\times10^{3}  E_{51}^{3/14} n_{sh}^{-4/7} {\rm\ yr},
\end{equation}
where $n_{sh}$ is the ambient gas density and $t_{\rm sh,20} =
t_{\rm shell}/(20\,000 {\rm\ yr}).$

\begin{figure}
\centering
\includegraphics[height=8.5cm]{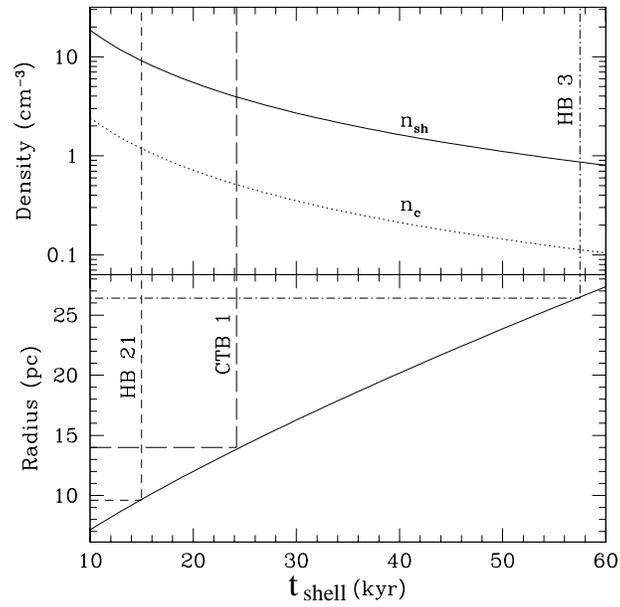}
\caption{Properties of a supernova remnant at the time of radiative
  shell formation. The horizontal axis gives the age at which shell
  formation occurs, and the vertical axes give the corresponding
  radius (lower panel) and density (upper panel). The radii of HB~21,
  CTB~1, and HB~3, as determined from their radio shells, are
  indicated, along with corresponding values for the shell ($n_{sh}$) and
  central ($n_c$) densities.  Note that uncertainties in the radius lead to uncertainties in the derived density and shell formation time, but these are not significant. }
\label{fig-rad-model}
\end{figure}

 Using a canonical SNR explosion energy $E_{51}=10^{51}$~ergs, we plot
$R_s$ and $n_{\rm sh}$ for a range of $t_{\rm shell}$ values in 
Figure~\ref{fig-rad-model}. Taking the observed shock radius for 
each remnant (as found from radio observations) to be the radius at
 which the SNR has entered the radiative phase we can set 
a lower limit to the ambient density. The resulting time required
 for the onset of the radiative phase at the current SNR radius (which is
an upper limit to the age, assuming the radiative phase was reached sometime
before the current epoch) is about two times higher than the age derived
 from the Sedov and evaporation cloud models for all three SNRs. 
 Note that uncertainties in the radius lead to uncertainties in the derived density and shell formation time, but these are not significant. 
 If thermal conduction is important in the SNR interior, then
 the radiative shell formation stage could be reached even earlier. More
 detailed modeling is required to determine whether thermal conduction
effects are fully consistent with the observations.

To estimate the central density $n_c$ of the X-ray
 emitting gas we use relations from \citet{cox99}:
\begin{equation}
n_c/n_{sh} = 0.334 (t_{\rm cool}/t)^{18/25} 
[(t/t_{\rm cool})^{1/25} - 0.44^{1/25}]^{18/71}
\label{eq-nc}
\end{equation}
 where $t_{\rm cool} \approx (7/6)t_{\rm shell}$ is the cooling time for 
the SNR. The calculated $n_c$ values are also plotted in 
Figure~\ref{fig-rad-model};
these represent lower limits for each indicated SNR. They are
significantly higher than the electron densities derived from the
emission measure in Table~\ref{tab-results} for HB~21 and CTB~1, but
comparable for HB~3.
 We note that equation~\ref{eq-nc} corresponds to full conduction, but
 it is weakly dependent on this; reduction of the conduction by a
 factor of ten reduces the central density by a factor of less then two.
Thus, even though the full conduction case is inconsistent with the
 assumptions made in deriving the onset of the radiative phase (because
the internal pressure is reduced in the presence of conduction), the
 associated central density predictions are not particularly
 sensitive to this. We conclude that the ambient densities required for the
SNRs discussed here to reach the radiative phase by the current ephoch are
consistent with inferred densities around the remnants, but that the central
densities derived by our spectral fits are lower than expected under such a
scenario. It is possible that the assumed volume for the X-ray emission
region has been overestimated in deriving these densities, but current
data are too sparse to better quantify the distribution of the multiple
emission components. Deeper X-ray observations of these SNRs are needed to
more fully investigate their evolutionary states.

\subsection{Presence of ejecta in MM SNRs}
\label{sec-ejecta}

The detection of the enhanced abundances in three MM SNRs studied here
 suggests that we detect contributions from shocked SN ejecta.
  Enhanced abundances have been found in a significant number of MM SNRs now.  
In Table~\ref{tab-mm} we list the SNRs identified as MM type, and briefly 
summarize their properties. Of 23 SNRs, 10 show evidence for enhanced abundances,
and 5 of those detections are confirmed by {\em Chandra} 
or {\em XMM} observations, including two SNRs studied here.
  Some of the remnants have multiple-component plasmas and enhanced
 abundances, like CTB~1, possibly indicating that the ejecta and swept-up components are
 spectrally resolved. In other SNRs with a single component plasma and
 enhanced abundances, like HB~21 and HB~3, the ejecta
 component might be more mixed with swept-up material and better
 statistics are needed to resolve the two components. We excluded
 W49B from the sample of MM SNRs with enhanced abundances  
because this remnant might be ejecta-dominated rather than swept-up
 mass dominated, and therefore not a true MM SNR \citep[for a different view see][]{kawasaki05}. However, W49B might
 prove to be in a pre-MM stage, while 3C~391, which has
 solar abundances, might represent a later stage of MM
 SNRs when ejecta is well mixed with swept-up mass. Kes~79 is another
 SNR classified as MM that has solar abundances \citep{sun04}. However, Kes~79 has
 a unique multiple-shell structure rather than diffuse or clumpy
 central brightness like other MM SNRs, and therefore might also not
 be a true MM SNR \citep{sun04}. We also note that 
W28 presently exhibits different properties from all of the other MM
SNRs. W28 has a cool shell in the north-west region where the SNR is
encountering a dense molecular cloud, and a hot plasma in the
south-west region where the SNR seems to be breaking out in a lower
density region \citep{rho02}. These two components have radio
counterparts, and are obviously related to the
large-scale inhomogeneities in the ambient medium that are influencing
the radio emission as well. However, it is the two-temperature thermal
 central emission of W28 that is most intriguing. 
Preliminary {\em Chandra} results
imply that soft emission is coming from a clumps embedded in hot
diffuse gas \citep{rho04}. 
   
It is often found that the evaporating cloud or thermal conduction models 
are able to explain individual MM cases, but there is general failure to
 accurately reproduce the central brightness profile, and 
a contribution from SN ejecta was suggested to play role in  
enhancing central brightness \citep[e.g.,][]{slane02}.  
 Recent results from {\em Chandra} observations of W44 show that 
 an elemental abundance gradient is present in this SNR, and that it 
has a significant contribution to the steepness of radial brightness profile 
\citep{shelton04}. Accurate determination of elemental abundances and their 
 distribution is therefore important in studying MM SNRs. 
HB~3 offers an interesting case because of the possibility of
under-abundant Fe. Under-abundant Fe, in contrast to enhanced abundances of Ne,
Si and S, have been found in W44. \citet{shelton04}  assign the Ne, 
Si and S enhancement to SN ejecta, and suggest that Fe originates 
from dust destruction and is under-abundant because Fe is still mostly
  condensed onto grains.


\section{CONCLUSIONS}

We present  archival  {\em ASCA} and {\em ROSAT} observations of three
mixed-morphology SNRs, HB~21, CTB~1 and HB~3, as well as archival {\em
Chandra} observations of CTB~1 and  archival {\em XMM-Newton}
observations of HB~3.  As found in many other MM SNRs, the morphology
of the SNRs is characterized by bright central emission and faint,
diffuse emission that extends up to the radio
boundary.  There is a good correspondence between the {\em ROSAT} and
{\em ASCA} features.  We confirm the thermal origin of the central
plasma  and detect enhanced abundances, which imply the presence of SN
ejecta in these SNRs.  HB~21 was found to have a single-component
plasma temperature of $\sim 0.6$~keV and enhanced abundances of Si 
and S. {\em ASCA} and {\em Chandra} data of CTB~1  require a
two-temperature thermal component, with a cooler component with $kT
\sim 0.2$~keV having solar abundances, and warmer component with $kT \sim
0.8$~keV having enhanced abundances of Mg. {\em ASCA} data from HB~3
require a $\sim 0.2$~keV thermal component with enhanced abundances of
Mg. {\em XMM} data from HB~3, however, allow for two possibilities ---
enhanced abundances of O and Mg, with column density and plasma
temperature consistent with {\em ASCA} results, or different column
density and plasma temperature values with either  enhanced abundance
of Ne or under-abundant Fe and solar O.  We revise the list of SNRs suggested to
be of MM type, and briefly examine some of their properties. A
significant number of MM SNRs have been found to have enhanced
abundances, which {\em Chandra} data of  W44 have shown to be crucial
for producing the steep radial  brightness profile in MM SNRs
\citep{shelton04}.

{\em ASCA} observations revealed  a compact source in the southern
part of HB~3. The column density value found towards the source
differs from that found towards the SNR and they are likely
unrelated. The spectrum is best described with a power law model with
photon index of $\sim 2.7$, or a blackbody model with temperature
of $\sim 2$~keV. The unabsorbed X-ray flux in the 0.5--10.0~keV band
is $\sim$\e{-12}\ergs.


\acknowledgements{}

We thank Manami Sasaki for help with {\em ROSAT} data reduction and
Randall Smith for helpful comments. This research has made 
use of data obtained through the High Energy
Astrophysics Science Archive Research Center Online Service, provided
by the NASA/Goddard Space Flight Center. 
We acknowledge the use of NASA's SkyView facility
 (http://skyview.gsfc.nasa.gov) located at NASA Goddard
 Space Flight Center. This work was supported in part by the National Aeronautics and 
Space Administration through contract NAS8-39073.

\clearpage

\begin{deluxetable}{lccccr}
\tablecaption{Archival {\em ASCA} data used in this study.
\label{tab-data-asca}}
\tabletypesize{\scriptsize}
\tablewidth{0pt}
\tablehead{\colhead{Source} & \colhead{Sequence} & \colhead{Observing} &
\colhead{GIS} & \colhead{SIS} & \colhead{SIS} \\
\colhead{ } & \colhead{number} & \colhead{date} &
\colhead{exposure} & \colhead{exposure} & \colhead{mode}}
\startdata

HB 21-P1  & 55053000 & 1997-06-09 & 39 ks & 35 ks & 1-CCD, bright \\
HB 21-P2  & 55054000 & 1997-06-10 & 38 ks & 34 ks & 1-CCD, bright \\
CTB 1-W   & 54026000 & 1996-01-21 & 58 ks & 36 ks & 2-CCD, bright  \\
CTB 1-NE  & 54027000 & 1996-01-22 & 42 ks & 37 ks & 2-CCD, bright  \\
HB 3-N1   & 54009000 & 1996-08-27 & 13 ks & 12 ks & 2-CCD, bright  \\
HB 3-N2   & 54009010 & 1996-08-28 & 30 ks & 23 ks & 2$\times$2-CCD\tablenotemark{a}, bright  \\
HB 3-N3   & 54009020 & 1996-08-29 & 33 ks & 29 ks & 2-CCD, bright  \\ 

\enddata
\tablenotetext{a}{complementary mode}
\end{deluxetable}


\begin{deluxetable}{lccc}
\tablecaption{Archival {\em ROSAT} data used in this study.
\label{tab-data-rosat}}
\tabletypesize{\scriptsize}
\tablewidth{0pt}
\tablehead{\colhead{Source} & \colhead{Sequence} & \colhead{Observing} &
\colhead{Exposure} \\
\colhead{ } & \colhead{number} & \colhead{date} &
\colhead{(s)} }
\startdata

HB 21 - CENTER & 500220n00 & 1998-07-23 & 4891 \\
HB 21 - NE     & 500221n00 & 1998-07-23 & 2768 \\
HB 21 - SE     & 500222n00 & 1998-07-23 & 3778 \\
HB 21 - SW     & 500223n00 & 1998-07-23 & 708 \\
HB 21 - NW     & 500224n00 & 1998-07-23 & 4442 \\ 
HB 21 - N      & 500225n00 & 1998-07-23 & 4370 \\

CTB 1 - FIELD 1 & 500155n00 & 1997-01-15 & 6291 \\
CTB 1 - FIELD 2 & 500156n00 & 1997-01-15 & 8940 \\

HB 3 - CENTER  & 500181n00 & 1996-10-23 & 2195 \\
HB 3 - N       & 500182n00 & 1996-10-23 & 3069 \\
HB 3 - SE      & 500183n00 & 1996-10-23 & 3906 \\
HB 3 - SW      & 500184n00 & 1996-10-23 & 2949 \\

\enddata
\end{deluxetable}

\clearpage

\begin{deluxetable}{lcccc}
\tabletypesize{\scriptsize}
\tablecolumns{5} 
\tablewidth{0pt} 
\tablecaption{Results from spectral models fits to {\em ASCA} data of 
HB~21, with the 90\% confidence ranges.  For comparison, we list $\chi_\nu^{2}$ values for the solar and enhanced abundances cases, as well as the statistical significance of the improved fit.\label{tab-fit-hb21}}
\tablehead{\colhead{ } & \colhead{P1-SIS} & \colhead{P1-whole} & 
\colhead{P2-SIS} & \colhead{P2-whole} \\
 \colhead{Parameter} & \colhead{(2SIS+2GIS)} & \colhead{(2GIS)} & 
\colhead{(2GIS+2SIS)} & \colhead{(2GIS)} \\
\colhead{ } & \colhead{VNEI} & \colhead{VNEI} & 
\colhead{VNEI} & \colhead{VNEI}
}
\startdata
$N_{\rm H}$ (\ee{21}~cm$^{-2}$) &  3.1$^{+0.3}_{-0.2}$ & 2.0$^{+0.5}_{-0.5}$ 
  & 4.0$^{+0.4}_{-0.2}$ & 3.0$^{+0.5}_{-0.5}$ \\
$kT$ (keV) & 0.63$^{+0.03}_{-0.02}$ & 0.58$^{+0.04}_{-0.03}$ & 
 0.68$^{+0.03}_{-0.02}$ & 0.64$^{+0.04}_{-0.02}$ \\
Si & 1.4$^{+0.2}_{-0.3}$ &  2.0$^{+0.4}_{-0.3}$ & 
     1.8$^{+0.3}_{-0.3}$ & 1.7$^{+0.3}_{-0.3}$\\
S & 1.0\tablenotemark{a} & 4.2$^{+2.1}_{-1.7}$ & 1.0 & 2.8$^{+1.3}_{-1.2}$\tablenotemark{a} \\
$n_e t$ (cm$^{-3}$\,s) & 3.7\ee{13}  ($\ge$3.4\ee{13}) & 4.8\ee{13} ($\ge$4.1\ee{13}) &
2.3\ee{13} ($\ge$2.0\ee{13}) & 1.6\ee{13}  ($\ge$1.3\ee{13}) \\ 
EM (cm$^{-3}$) & 1.2\ee{55} & 8.8\ee{55} & 1.1\ee{55} & 3.3\ee{55}\\
$\chi_\nu^{2}$/dof (solar) & 1.38/236 & 1.07/302 & 1.29/232 & 1.07/337 \\
$\chi_\nu^{2}$/dof (abund) & 1.36/235 & 0.97/300 & 1.21/231 & 1.01/335 \\
improvement & $2\sigma$ & $>4\sigma$ & $>3\sigma$ & $>4\sigma$ \\
\enddata
\tablenotetext{a}{The elemental abundance has been fixed at solar value.}
\end{deluxetable}


\begin{deluxetable}{lcccc}
\tabletypesize{\scriptsize}
\tablecaption{Results from spectral models fits to {\em
ASCA} data of CTB~1, with the 90\% confidence ranges.  For comparison, we list $\chi_\nu^{2}$ values for the solar and enhanced abundances case, as well as the statistical significance of the improved fit.\label{tab-fit-ctb1}}
\tablewidth{0pt}
\tablehead{
\colhead{ } & \colhead{W-SIS} & \colhead{W-whole} & \colhead{NE-whole}
& \colhead{center} \\
\colhead{Parameter} & \colhead{(2GIS+2SIS)} & \colhead{(2GIS)} & \colhead{(2GIS)} & \colhead{(4ACIS)} \\
 \colhead{center} & \colhead{VRAY+VRAY } & \colhead{VRAY} & \colhead{VRAY} & \colhead{VNEI+VNEI}
}
\startdata
$N_{\rm H}$ (\ee{21}~cm$^{-2}$) &  5.2$^{+0.8}_{-2.0}$ & 7.5$^{+0.8}_{-0.7}$ & 8.7$^{+0.7}_{-0.5}$ & 
   5.7$^{+0.7}_{-0.7}$ \\
$kT_1$ (keV) & 0.19$^{+0.09}_{-0.03}$ & 0.26$^{+0.03}_{-0.03}$ &   0.18$^{+0.00}_{-0.01}$ &
   0.20$^{+0.04}_{-0.01}$ \\
Mg$_1$ & 1.0\tablenotemark{a} & 1.0 & 1.0 & 1.0 \\
$n_e t_1$ (cm$^{-3}$\,s) & CIE & CIE  & CIE & 8.9\ee{12} ($\ge$1.5\e{12}) \\ 
EM$_1$ (cm$^{-3}$) & 9.9\ee{56} &  4.6\ee{57} & 2.8\ee{58} & 2.0\ee{56} \\
$kT_2$  &  0.82$^{+0.09}_{-0.06}$ & --- & --- & 0.86$^{+0.03}_{-0.06}$  \\
Mg$_2$ & 2.7$^{+0.9}_{-0.5}$ & 1.0 & 1.0 & 3.1$^{+1.0}_{-0.4}$ \\
$n_e t_2$ (cm$^{-3}$\,s) &  CIE & ---  & --- & 1.4\ee{13} ($\ge$1.1\ee{12}) \\ 
EM$_2$  & 9.6\ee{55} & --- & --- &  1.9\ee{55}  \\
$\chi_\nu^{2}$/dof (solar) & 1.38/203 & 1.15/ 248 & 1.14/210 & 0.87/469 \\
$\chi_\nu^{2}$/dof (abund) & 1.29/202 & --- & --- & 0.80/468 \\
improvement & $>3\sigma$ & --- & --- & $>5\sigma$ \\
\enddata
\tablenotetext{a}{The elemental abundance has been fixed at solar value.}
\end{deluxetable}


\begin{deluxetable}{lccccc}
\tablecaption{Results from spectral models fits to {\em
ASCA} data of HB 3, with the 90\% confidence ranges.  For comparison, we 
list $\chi_\nu^{2}$ values for the solar and enhanced abundances case, as well as the statistical significance of the improved fit.
\label{tab-fit-hb3}}
\tabletypesize{\scriptsize}
\tablewidth{0pt}
\tablehead{
\colhead{ } & \colhead{N2-SIS0} & \colhead{N2-SIS1}  & \colhead{N3-north} & \colhead{N3-south} &
\colhead{XMM-center\tablenotemark{a}}  \\
 \colhead{Parameter} & \colhead{(2GIS+SIS)} & \colhead{(2GIS+SIS)} 
 & \colhead{(2GIS+2SIS)} & \colhead{(2GIS+2SIS)} & \colhead{(pn)}\\
\colhead{ } & \colhead{VRAY} & \colhead{VRAY}  & \colhead{VRAY} & \colhead{VRAY} &
\colhead{VRAY}  
}
\startdata
$N_{\rm H}$ (\ee{21}~cm$^{-2}$) & 8.6$^{+0.3}_{-0.3}$ & 8.8$^{+0.4}_{-0.3}$ 
  & 7.0$^{+1.8}_{-0.7}$ & 6.5$^{+1.0}_{-1.4}$  &  9.2$^{+0.5}_{-0.6}$ \\
$kT$ (keV) & 0.18$^{+0.02}_{-0.01}$ & 0.18$^{+0.02}_{-0.01}$ & 
 0.27$^{+0.03}_{-0.05}$ & 0.29$^{+0.07}_{-0.04}$ &  0.18$^{+0.02}_{-0.03}$ \\
O & 1.0\tablenotemark{b} & 1.0 & 1.0 & 1.0 & 1.9$^{+0.4}_{-0.4}$ \\
Mg & 2.6$^{+0.6}_{-0.5}$ & 1.6$^{+0.6}_{-0.3}$\tablenotemark{c} & 1.0
& 1,0  &   2.0$^{+0.7}_{-0.6}$  \\
EM (cm$^{-3}$) & 1.1\ee{58} & 1.4\ee{58} & 5.5\ee{56} & 3.9\ee{56} & 5.5\ee{57} \\
$\chi_\nu^{2}$/dof (solar) & 1.56/274 & 1.60/280 &  1.25/144 & 1.36/188 & 1.22/2066 \\
$\chi_\nu^{2}$/dof (abund) & 1.42/273 & 1.56/279 &  --- & --- & 1.19/2064 \\
improvement & $>4\sigma$ & $>3\sigma$ & --- & --- &  $>5\sigma$ \\
\enddata
\tablenotetext{a}{see section \S\ref{sec-xmm-spec} for additional comments}
\tablenotetext{b}{The elemental abundance has been fixed at solar value.}
\tablenotetext{c}{fit was not improved by thawing this elemental abundance}
\end{deluxetable}


\begin{deluxetable}{lccc}
\tabletypesize{\scriptsize}
\tablecolumns{4} 
\tablewidth{0pt} 
\tablecaption{SNR parameters derived from spectral analysis using
     the central SNR regions and different SNR evolutionary models. \label{tab-results}}
\tablehead{
\colhead{ } & \colhead{HB~21} & \colhead{CTB~1} &  \colhead{HB~3} \\
\colhead{Parameter} &  \colhead{P2-SIS} & \colhead{W-whole}  & \colhead{N2-SIS0} }
\startdata
R (arcmin) & 40 & 15 & 40 \\
R (pc) & 9.6 & 14.0 & 26.4 \\
d (kpc) & 0.8 & 3.1 & 2.2 \\
V (cm$^{-3}$) & 2\ee{57} & 5\ee{57}  & 8\ee{58} \\
\cutinhead{Sedov model (adiabatic stage)\tablenotemark{a}} 
$n_H$ (cm$^{-3}$) &  $0.07\pm0.01$ & $0.15\pm0.04$ & $0.32\pm0.10$ \\
$kT_s$ (keV) & $0.53\pm0.03$ & $0.20\pm0.02$  & $0.14\pm0.02$  \\
$v_{s}$ (\kms) & $670\pm33$ & $410\pm45$ & $340\pm37$ \\
$t_{\rm SNR}$ (yr) & $5600\pm280$ & $13100\pm1440$  & $30000\pm3300$  \\
$E_{\rm SNR}$ (erg) & ($1.3\pm0.3$)\ee{49} & ($3.4\pm1.2$)\ee{49} 
& ($3.4\pm1.5$)\ee{50}  \\
$M$ (\msol) & $0.16\pm0.04$ & $31\pm11$ & $31\pm14$ \\
\cutinhead{Evaporating clouds\tablenotemark{b}} 
$kT_s$ (keV) & 0.57--0.73 & 0.21--0.27 & 0.14--0.18 \\
$v_s$ (\kms) & 700--790 & 420--480 & 350--400 \\
$t_{\rm SNR}$ (yr) & 5400--4800 & 13000--11500 & 30000--25000 \\
$E_{\rm SNR}$ (erg) &  (3--8)\ee{50} & (8--20)\ee{49} & (7--20)\ee{50}  \\
\enddata
\tablenotetext{a}{calculated for best-fit values from
  Tables~\ref{tab-fit-hb21}~to~\ref{tab-fit-hb3} }
\tablenotetext{b}{calculated for range: $K/q=1.0-1.3$ (see
  \S\ref{sec-evap}) }
\end{deluxetable}


\begin{deluxetable}{lcccc}
\tabletypesize{\scriptsize}
\tablecolumns{5} 
\tablewidth{0pt} 
\tablecaption{A list of MM SNRs and some of their properties; the first part of the
  table lists the SNRs studied by \citet{rho98}, and the second part lists
  other SNRs suggested to be of MM type. \label{tab-mm}
}
\tablehead{ \colhead{SNR Name}  & \colhead{Enhanced} &
  \colhead{Multiple} & \colhead{Molecular} & \colhead{OH(1720~MHz)} \\
\colhead{} & \colhead{abundances\tablenotemark{a}} & 
\colhead{components} &\colhead{material} & \colhead{masers\tablenotemark{b}} 
}
\startdata
G6.4--0.1 (W28)       & N (ASCA) [1] & Y [1] & Y [2]  & Y [3]  \\
G31.9+0.0 (3C 391)    & N (ACIS) [4]  & N [4]  & Y [5]  & Y [6] \\
G33.6+0.1 (Kes 79)    & N (ACIS) [7]  & N [7]  & Y [8]  & N [6] \\
G34.7--0.4 (W44)      & Y (ACIS) [9] & N [9] & Y [10]  & Y [11] \\
G41.1--0.3 (3C 397)   & Y (ACIS) [12] & Y [12] & ? & N [13] \\
{\em G43.3--0.2 (W49B)}\tablenotemark{c}  & {\em ejecta?} [14] & --- & --- & ---\\
G49.2--0.7 (W51C)     &  N (ASCA) [15] & Y [15] & Y [16] & Y [13] \\
G53.6--2.2 (3C 400.2) &  N (ASCA) [17] & N [17] & ? & ? \\
G82.2+5.3 (W63)       & Y (ROSAT+ASCA) [18] & Y [18] & ? & ? \\
G89.0+4.7 (HB21)      & Y (ASCA) [19] & N [19] & Y [20] & N [6] \\
G93.7--0.2 (CTB 104A) &  N (ROSAT) [21] & N [21] & ? & N [6] \\ 
G116.9+0.2 (CTB 1)    & Y  (ASCA, ACIS) [19] & Y [189] & ? & N [6] \\
{\em G119.5+10.2 (CTA~1)}\tablenotemark{d} & {\em PWN} [22] & --- & --- & --- \\ 
G132.7+1.3 (HB~3)      & Y (ASCA, pn) [19] & N [19] & Y [23] & N [6]\\
G189.1+3.0 (IC 443)   & N (ASCA) [24] & Y [24] & Y [25] & Y [11]\\
G290.1--0.8 (MSH 11--61$A$) & Y (ASCA) [26] & N [26] & ? & N [6]\\
{\em G327.1--1.1}\tablenotemark{d}  &  {\em PWN} [27] & --- & --- & --- \\ 
G327.4+0.4 (Kes 27)   &  N (ASCA) [28] & N [28] & ? &  N (ATCA) [13]\\
\cutinhead{ }
Sgr A East (G0.0+0.0)     & Y (ACIS) [29] & N [29] & ? & Y [30] \\
G65.3+5.7                 & N (ROSAT) [31] & N [31] & ? & ? \\
G156.2+5.7                & Y (ASCA) [32] & N [32] & ? & ? \\
G160.9+2.6 (HB9)          & N (ROSAT) [33] & N [33] & ? & N [6] \\
G166.0+4.3 (VRO 42.05.01) & N (ASCA) [34] & N [34] & ? & N [6]\\
G272.2--3.2               & N (ASCA) [35] & N [35] & ? & ?\\
G357.1-0.1 (Tornado)      & N (ACIS) [36] & N [36] & Y [37] & Y [38] \\
G359.1--0.5               & Y (ASCA) [39] & Y [39] & Y [40] & Y [38] \\
\enddata

\tablenotetext{a}{indicates at the sensitivity observed with telescope
  noted in parenthesis}

\tablenotetext{b}{non-detection with a single-dish telescope, unless
  telescope given in parenthesis}

\tablenotetext{c}{the central emission might be ejecta-dominated, so the SNR might 
not belong to the MM class}

\tablenotetext{d}{the central emission subsequently found to 
be dominated by non-thermal 
emission, so the SNRs do not belong to the MM class anymore}

\tablerefs{ [1] \citet{rho02}; [2] \citet{reach05}; [3] \citet{frail94};
 [4] \citet{chen04}; [5] \citet{reach99}; [6] \citet{frail96};
 [7] \citet{sun04}; [8] \citet{green92}; [9] \citet{shelton04}; 
 [10] \citet{frail98}; [11] \citet{claussen97}; [12] \citet{safi04};
 [13] \citet{green97};   
[14] \citet{hwang00}; [15] \citet{koo02}; [16] \citet{koo97};
  [17] \citet{yoshita01}; [18] \citet{mavro04}; [19] this work;
  [20] \citet{koo01}; [21] \citet{rho98}; [22]  \citet{slane97};
  [23] \citet{huang86}; [24] \citet{kawasaki02}; [25] \citet{vand93};
  [26] \citet{slane02}; [27] \citet{bocchino03}; [28] \citet{enoguchi02}; 
  [29] \citet{maeda02};  [30] \citet{yusef96}; [31] \citet{shelton04-2};
  [32] \citet{yamauchi99}; [33] \citet{leahy95}; [34] \citet{guo97};
  [35] \citet{harrus01}; [36] \citet{gaensler03}; [37] \citet{lazendic04};
   [38] \citet{yusef99}; [39] \citet{bamba00}; [40] \citet{lazendic02}; }
\end{deluxetable}


\clearpage

\clearpage

\end{document}